\documentstyle[12pt,titlepage]{article}
\input epsfig.sty 

\renewcommand{\theequation}{\thesection.\arabic{equation}}

\font\grande=cmr10 scaled \magstep4
\font\medio=cmr10 scaled \magstep2
\outer\def\beginsection#1\par{\medbreak\bigskip
      \message{#1}\leftline{\bf#1}\nobreak\medskip\vskip-\parskip
      \noindent}

\def\laq{\raise 0.4ex\hbox{$<$}\kern -0.8em\lower 0.62 ex\hbox{$\sim$}}
\def\gaq{\raise 0.4ex\hbox{$>$}\kern -0.7em\lower 0.62 ex\hbox{$\sim$}}

\begin{document}
\bibliographystyle{unsrt}
\titlepage
\begin{flushright}
CERN-TH-97-264\\
DAMTP-97-108\\
\end{flushright}
\begin{center}
{\grande Primordial Hypermagnetic Fields }\\
\vspace{5mm}
{\grande and Triangle Anomaly}\\
\vspace{15mm}
\centerline{M. Giovannini$^{(b)}$ and 
M. E. Shaposhnikov$^{(a)}$\footnote{Electronic addresses: 
mshaposh@nxth04.cern.ch,~~m.giovannini@damtp.cam.ac.uk}}
\bigskip
\centerline{$^{(a)}${\em CERN, Theory Division, CH-1211 Geneva 23,
Switzerland}}
\smallskip
\centerline{$^{(b)}${\em DAMTP, Silver Street, CB3 9EW Cambridge, 
United Kingdom}}
\end{center}
\vspace{10mm}
\centerline{\medio  Abstract}
\noindent
The high-temperature plasma above the electroweak scale $\sim 100$ GeV
may have contained a primordial hypercharge magnetic field whose
anomalous coupling to the fermions induces a 
transformation of the hypermagnetic energy density into  fermionic
number. In order to describe this process, we generalize the ordinary
magnetohydrodynamical equations to the anomalous case. We show that
 a  not completely homogeneous hypermagnetic background  induces
fermion number fluctuations, which can be  expressed in terms 
of a generic hypermagnetic field configuration. 
We  argue  that, depending upon the various particle physics
parameters involved in our estimate
 (electron Yukawa coupling, strength of the electroweak  phase
transition) and upon the hypermagnetic energy spectrum,
 sizeable  matter--antimatter fluctuations can be generated in the
plasma. These fluctuations may modify the predictions of the standard
Big Bang nucleosynthesis (BBN). We derive constraints on the magnetic fields
from the requirement that the homogeneous BBN is not changed. We
analyse the influence of primordial magnetic fields on the electroweak
phase transition and  show that some specific configurations of the
magnetic field may be converted into net baryon number at the
electroweak scale.

\newpage
\renewcommand{\theequation}{1.\arabic{equation}}
\setcounter{equation}{0}
\section{Introduction} 

There are no compelling reasons why magnetic
fields should not have been present in the  early Universe.
Moreover, it can be argued that the existence of some magnetic fields at
high temperatures is a quite natural phenomenon. Indeed,
the presence of large scales magnetic fields in our observed Universe is a
well established experimental fact. Since their first evidence in
diffuse astrophysical plasmas beyond the solar corona \cite{1,2},
magnetic fields have been detected in our galaxy and in our local
group through Zeeman splitting and through Faraday rotation
measurements of linearly polarized radio waves.
The Milky Way possesses a magnetic field whose strength is of the order
of the microgauss corresponding to an energy density roughly
comparable with the energy density today stored in the Cosmic
Microwave Background Radiation (CMBR) energy spectrum peaked 
around a frequency of $30$ GHz.
Faraday rotation measurements of radio waves from extra-galactic
sources also suggest that various spiral galaxies are  endowed with 
magnetic fields whose intensities are of the same order as that of
the Milky Way \cite{3}.
The existence of magnetic fields at even larger scales (intergalactic
scale, present horizon scale, etc.) cannot be excluded, but it is
still quite debatable since, in principle, dispersion measurements 
(which estimate the electron density along the line of sight) cannot be
applied in the intergalactic medium because of the absence of pulsar
signals \cite{3}.

If the existing galactic magnetic field is naively blue-shifted to 
earlier epochs, one then finds that the Universe was always filled by a
magnetic field of a considerable amplitude, with energy density of the
order of the energy density of $\gamma$ quanta. Of course, this
consideration does not take into account different physical 
processes operating at the galactic  
scale, such as the dynamo mechanism 
\cite{2,6} or the anisotropic collapse mechanism \cite{7}, which
change considerably the naive scaling law $|\vec{H}|\sim T^2$. In any case,
it is widely believed that some seed fields of primordial origin are
necessary for the successful generation of the galactic magnetic 
fields \cite{5}.

Looking at this problem from a more theoretical 
 side, there are several mechanisms
that may successfully 
generate large enough magnetic seeds coherent on different scales. 
Magnetic seeds can be produced either during a first-order
quark--hadron phase transition  \cite{8,9} or during 
the electroweak phase transition \cite{5c,10,11,12}.
Recently it was also suggested that a primordial asymmetry encoded in
the right electron number can be converted in a quite large
hypercharge seed during the symmetric phase of the electroweak 
theory \cite{14}.
The seeds could also be the result of the parametric amplification of
the quantum mechanical (vacuum) fluctuations of some primordial gauge
field, in the same way as in General Relativity the quantum mechanical
fluctuations of the tensor modes of the metric can be amplified,
producing, ultimately, a stochastic gravity-waves background \cite{15}.
The essential ingredient of the large-scale magnetic field generation
is the breaking of conformal invariance in the coupling of the
electromagnetic field to gravity \cite{16,17}.
Reasonable
seeds could also be produced if the inflaton is coupled to the Maxwell
term in a chaotic inflationary scenario \cite{18}.
In the string theory low-energy effective action,  the 
dilaton field provides a unified value of the gravitational and gauge 
coupling at the string scale  and it naturally breaks the scale
invariance of the electromagnetic and gauge couplings (also in 
four dimensions) without providing a gravitational mass for the
photon. Sizeable seeds, coherent over the galactic scale, can be
generated \cite{19,19b}.

The possible existence of magnetic fields in the early Universe has a
number of interesting cosmological implications.
For example, magnetic fields at small scales may influence the Big
Bang nucleosynthesis (BBN) and change the primordial abundances of
the light elements \cite{20} by changing the rate of the Universe
expansion at the corresponding time. The success of the
standard BBN  scenario can provide an interesting
set of bounds on the intensity of the magnetic fields at that epoch \cite{20}.

Long-range stochastic magnetic fields that
 possibly existed at the decoupling epoch
might have induced anisotropies in  the microwave 
sky \cite{19b}. 
The existence of a completely homogeneous field coherent over the
horizon at the present  epoch can be interestingly constrained by
the COBE observations \cite{64,66}.
Conversely, if the CMBR is linearly polarized,
its polarization plane could have been rotated by the presence of a
sufficiently energetic magnetic field coherent over the horizon size
at the decoupling epoch \cite{68}. 
Faraday rotation measurements applied to the galactic (synchrotron)
emission can also provide interesting constraints \cite{65} on large-scale
magnetic fields (even though these are coherent over scales smaller than the
present horizon). 

In this paper we address the question of whether there can be any
cosmological consequences from the fact that
 magnetic fields existed prior to the 
electroweak phase transition, when the background temperature was
 $T>T_{c} \sim 100$ GeV. At these
temperatures the electroweak symmetry is restored and the magnetic
field is replaced by the hypermagnetic one.  The hypercharge  
field, unlike the ordinary magnetic field, 
has an anomalous coupling to the  fermions.
This fact will play a crucial role in our considerations.
  The origin of primordial hypermagnetic fields is not
essential for us and consequently we simply assume that they were
generated by some mechanism before the electroweak (EW) phase transition. 

We will show that, depending on the particle physics model and on the initial
spectrum of the  primordial magnetic fields, quite large fluctuations of
the baryon and lepton numbers may be generated. These fluctuations can
survive until the onset of BBN and create  unusual initial conditions
for the calculation of the light element abundances.
In particular a natural outcome of our considerations is a 
non-uniform distribution of baryon number, not necessarily
positive-definite. Matter--antimatter domains are then possible.
 The requirement that these
fluctuations are small enough in order not to conflict with the predictions of
the standard homogeneous BBN allows us to put quite a strong constraints on
the spectrum of the primordial magnetic fields.
Moreover we will argue that the primordial 
magnetic field may change the nature 
of the electroweak phase transition. Finally,  the existence of
primordial fields with  some specific topological structure may result in the
production of the net baryon number at the electroweak scale.  

The plan of the paper is the following.
In  Section 2 we derive our basic equations.
For an ordinary electromagnetic plasma, it is fairly well established
that the evolution of the magnetic fields can be described using the
magnetohydrodynamical (MHD) equations \cite{2,6,25b}. 
In the case of a high-temperature electroweak plasma the MHD
equations have to be generalized by taking into
account anomaly effects. In the new AMHD equations (anomalous MHD equations) 
the magnetic hypercharge fields turn out to be coupled to the
fermionic number density. As a consequence,  
the evolution equations of the anomalous charge densities
acquire a magnetic source term. In Section 3 we describe an
approximate solution of AMHD equations. We show that anomalous 
coupling gives  rise to instabilities, allowing the conversion of the
energy sitting in the fermionic degrees of freedom into magnetic
hypercharge fields and vice versa. This phenomenon is completely absent in
ordinary MHD. The presence of the fermionic number density 
produces a kind of ``ohmic'' current, which is parallel to the magnetic
hypercharge field. Also this phenomenon is quite new if compared with
the ordinary MHD equations (though something vaguely similar can
happen in the context of the dynamo mechanism \cite{2} in a parity-
non-invariant turbulent fluid).
In Section 4 we will apply our results to the case of 
stochastic hypercharge field
backgrounds, whereas in Section 5 we focus our attention on 
the possible phenomenological implications of our considerations for BBN.
Different bounds on primordial magnetic fields will also be analysed
in the light of our results. We will also discuss the dependence of
the  magnitude of the  baryon number fluctuations upon various
particle physics parameters, which can appear in the extensions of the
minimal standard model (MSM). 
In Section 6 we discuss the influence of the primordial hypermagnetic
fields on the dynamics of the 
 electroweak phase transition, and we will show how some specific
configurations of the hypermagnetic fields may create a net non-zero
amount of baryons. Section 7 contains our concluding remarks.

Part of the results of this paper has been already presented (in a more
compact form) in \cite{us}, (see also the closely related paper \cite{14}
where transformation of a finite number density of right electrons
into hypermagnetic fields has been considered). 

\renewcommand{\theequation}{2.\arabic{equation}}
\setcounter{equation}{0}
\section{Basic equations}
\subsection{Preliminaries}

Let us start from some qualitative considerations. A unique property of
``unbroken" U(1) gauge interaction is the absence of mass of its
corresponding vector particle. Static ``magnetic" fields are never
screened (in the absence of monopoles) and thus homogeneous fields can
survive in the plasma for infinitely long  times. On the contrary, electric
fields quickly decay because of the finite plasma conductivity $\sigma$
within a time scale $\sim 1/\sigma$. Then long-ranged non-Abelian
magnetic fields (corresponding to, e.g. the colour SU(3) or weak SU(2)
groups) cannot exist because at high temperatures the
non-Abelian interactions induce a ``magnetic'' mass gap $\sim g^2 T$.    
Also the non-Abelian electric fields decay because of the finite value
of the conductivity as it occurs for Abelian electric fields.
 Therefore, the only long scale field that  can
exist in the plasma for enough time must be associated with some
Abelian U(1) group. This statement, valid to all orders in
perturbation theory, has been confirmed non-perturbatively for the
electroweak theory by recent lattice studies in \cite{37}.
Under normal conditions (i.e.
small temperatures and small densities of the different fermionic
charges) the SU(2)$\times$U(1)$_Y$ symmetry is ``broken" down to
U(1)$_{EM}$, the massless field corresponding to U(1)$_{EM}$ is the
ordinary photon and the only long-lived field  in the
plasma is the ordinary magnetic one. At sufficiently high
temperatures, $T > T_c$, the SU(2)$\times$U(1)$_Y$ symmetry is
``restored", and non-screened vector modes $Y_\mu$ correspond to the
U(1)$_Y$ hypercharge group. Hence, if primordial fields existed at $T >
T_c$, they did correspond to hypercharge rather than to U(1)$_{EM}$. 

There are
essential differences between the interactions of magnetic fields and
the ones of hypermagnetic fields with matter. 
The ordinary electromagnetic field
has a vector-like coupling to the fermions, while the coupling of the
hypercharge fields is chiral. Thus, if hyperelectric ($\vec{{\cal
E}}_{Y}$) and hypermagnetic ($\vec{{\cal H}}_{Y}$) fields are present
simultaneously, they cause a variation of the fermionic number
according to the anomaly equation, $\partial_\mu j_\mu \sim
\frac{g'^2}{4\pi^2} \vec{{\cal H}}_{Y}\cdot \vec{{\cal E}}_{Y}$ (here
$g'$ the hypercharge gauge coupling constant). Now, the presence of
{\em non-homogeneous} hypermagnetic fields in the EW plasma with 
(hyper)conductivity $\sigma_c$ always implies the existence of a related
electric field, $\vec{{\cal E}}_{Y}\sim \frac{1}{\sigma_c} \vec{\nabla}
\times \vec{{\cal H}}_{Y}$. Since for a general stochastic magnetic
background $\langle(\vec{{\cal H}}_{Y}\cdot \vec{\nabla}\times
\vec{{\cal H}}_{Y})^2\rangle \neq 0$, the non-uniform hypermagnetic
field may absorb or release fermions and produce, ultimately, 
baryon and lepton density perturbations because of the anomaly equation. 

The behaviour of {\em cold} fermionic matter with non-zero anomalous Abelian
charges was considered in \cite{27} where it  was pointed out  that the
anomalous fermionic matter is unstable against the creation of 
Abelian gauge field
with non-zero Chern-Simons number, which eats up fermions because of
the anomaly. It was suggested in \cite{14} that the right electron number
density may be converted to the hypercharge field because of a
similar effect. The main aim of this paper is the study of the
opposite situation, namely we want to understand how hypercharge
fields may be converted into fermions in a {\em hot} environment.

\subsection{MHD equations for ordinary plasmas}

During the symmetric phase of the electroweak theory the evolution of
the background geometry is dominated by radiation.
The first assumption we will make is that prior to $T_{c} \simeq 100~{\rm
GeV}$
the geometry may be described by a conformally flat metric of the FRW type,
whose line element is:
\begin{equation}
ds^2=g_{\alpha\beta} dx^{\alpha} dx^{\beta}= 
a(\tau)^2(d\tau^2 - d \vec{x}^2), ~~~a(\tau)\sim \tau~~~
\label{2.1}
\end{equation} 
($\tau$ is the conformal time coordinate related to the cosmic time
coordinate $t$ as $dt= a(\tau) d\tau$).
We will also assume that the radiation-dominated stage started much
before the electroweak epoch at temperatures $T> T_{c}$.

The Weyl invariance of the ordinary Maxwell  equations in a
conformally flat FRW background geometry implies that the MHD
equations  in the
metric (\ref{2.1}) can be written \cite{40} as:
\begin{eqnarray}
\frac{\partial{\vec{H}}}{\partial\tau} = -\vec{\nabla}
\times \vec{{E}} 
,~~~~~
\frac{\partial{\vec{E}}}{\partial\tau}+ \vec{J} =
{\vec{\nabla}}\times \vec{H}, 
\nonumber\\
{\vec{\nabla}}\cdot{\vec{H}}=0,~~~~~~~~ ~~~~~~~~~~~~
{\vec{\nabla}}\cdot {\vec{E}}=0,
\nonumber\\
{\vec{\nabla}}\cdot\vec{J}=0,~~~ ~~~~
\vec{J}=\sigma ({\vec{E}} + \vec{v}\times{\vec{H}})
\label{2.3}
\end{eqnarray}
(${\vec{H}}=a^2 \vec{{\cal H }}$, $\vec{E}=a^2
\vec{{\cal E}}$; $\vec{A}= a \vec{\cal A}$; $\vec{J}=a^3 \vec{j}$; $\sigma=
\sigma_{c} a$; $\vec{{\cal H}}= \vec{\nabla}\times \vec{{\cal A}}$;
$\vec{{\cal E}} = - \frac{\partial}{\partial t}\vec{\cal A}$;  $\vec{{\cal
H}}$, $\vec{{\cal E}}$, $\vec{{\cal A}}$,
$\vec{j}$, $\sigma_{c}$ are  the flat-space quantities whereas
$\vec{H}$, $\vec{E}$, $\vec{A}$, $\vec{J}$, $\sigma$ are the
curved-space ones; $\vec{v}$ is the bulk velocity of the plasma).
We assumed that the plasma is locally electrically neutral
(${\vec{\nabla}}\cdot {\vec{E}}=0 $) over  length scales 
larger than the Debye radius.
We notice that the spatial gradients used in Eq. (\ref{2.3}) are
defined according to the metric (\ref{2.1}).

There are several approximations in which the above equations can be
studied. One is the so-called ideal (or superconducting) approximation
and the other is the real (or resistive) case \cite{25b,40b}.

In the ideal case $\sigma^{-1}=0$ and, from Ohm's law we can
immediately deduce that the electric field is orthogonal to the
magnetic one and it is also orthogonal to the bulk velocity of the
plasma:
\begin{equation}
{\vec{E}}\simeq - {\vec{v}}\times{\vec{H}}.
\label{2.4}
\end{equation}
Two important theorems of the ideal MHD, which follow from 
Eq. (\ref{2.4}), are the conservation of the
magnetic flux:
\begin{equation} 
\Phi = \int_{\Sigma}  \vec{H}\cdot d\vec{\Sigma}
\end{equation}
and of the magnetic helicity (Chern-Simons number) \cite{2,40b}:
\begin{equation}
N_{CS} = \frac{\alpha_{em}}{\pi}\int_{V} d^{3}x~ \vec{H}\cdot
\vec{A},~~~ \alpha_{em} = \frac{e^2}{4\pi},
\end{equation}
where $d\overline{\Sigma}$ is a closed surface in the plasma;
the volume integral is performed over a magnetic flux tube. 

If, on the contrary, the effect of the finite value of the
 conductivity is taken into account ($\sigma^{-1} \ll 1$) and the
 resistive Ohm law is employed, then  the induced  
electric field is not exactly
 orthogonal to the magnetic one and the conservation laws of the ideal
 MHD are corrected (in the  resistive approximation) by an expansion
 in powers of the resistivity, which can be explicitly computed and
 which we report at the lowest order in $\sigma^{-1}$:
\[
\frac{d}{d\tau}\Phi =
-\frac{1}{\sigma}\int_{\Sigma}{\vec\nabla}\times({\vec\nabla}
\times\vec{H})\cdot d\vec{\Sigma},
\]
\begin{equation}
\frac{d}{d\tau} N_{CS} =
-\frac{\alpha_{em}}{\pi \sigma}\int_{V} d^3 x ~\vec{H}\cdot\vec{\nabla}
\times\vec{H}.
\label{2.5}
\end{equation}
According to Eqs. (\ref{2.5}) the magnetic flux lines
evolve glued to
the hypercharged plasma element; also the sum of the link and twist number
of the magnetic flux tubes is always the same all  along the time
evolution, only provided that $\sigma^{-1}= 0$. 

The same dynamical information encoded in the magnetic flux
conservation is also contained in the magnetic diffusivity equation,
which can be derived according to Eq. (\ref{2.3})
\begin{equation}
\frac{\partial{\vec{H}}}{\partial\tau} =
{\vec{\nabla}}\times(\vec{v}\times{\vec{H}}) +
\frac{1}{\sigma}\nabla^2{\vec{H}}.
\label{2.7}
\end{equation}
The ratio of the two terms on the r.h.s. defines the 
 magnetic Reynolds number 
\begin{equation}
R \simeq
\frac{\sigma|\overline{\nabla}\times\vec{v}\times\vec{H}|}{|\nabla^2
\vec{H}|}\simeq v~l~\sigma.
\label{2.8}
\end{equation}
If $R\ll 1$ (for a given length scale $l$) the flux lines of the
magnetic field will diffuse through the plasma. If $R\gg 1$ the flux
lines of the magnetic field will be frozen into the plasma element.
From the magnetic diffusivity equation (\ref{2.5}) it is possible to
derive the typical structure of the dynamo term by carefully averaging
over the velocity field according to the procedure outlined in  \cite{2,6}.
By assuming that the motion of the  fluid is random and has zero mean
velocity, it is possible to average over the ensemble of the possible
velocity fields.
In more physical terms this averaging procedure of Eq. (\ref{2.7}) is
equivalent to averaging over scales and times exceeding the
characteristic correlation scale and time $\tau_{0}$ of the velocity
field. This procedure assumes that the correlation scale of the
magnetic field is much larger than the correlation scale of the
velocity field.
In this approximation the magnetic diffusivity equation can be written
as:
\begin{equation}
\frac{\partial\vec{H}}{\partial\tau} =
\alpha(\vec{\nabla}\times\vec{H}) +
\frac{1}{\sigma}\nabla^2\vec{H} 
\label{2.9}
\end{equation}
($\alpha 
= -\frac{\tau_{0}}{3}\langle\vec{v}\cdot\vec{\nabla}
\times\vec{v}\rangle$ is the so-called dynamo term, which vanishes
in the absence of vorticity; in this equation $\vec{H}$ is
the magnetic field averaged 
over times larger than $\tau_{0}$, which is the typical correlation
time of the velocity field). We can clearly see that the crucial
requirement for the all averaging procedure we described is that the
turbulent velocity field has to be ``globally'' non-mirror-symmetric.
It is  interesting to point out \cite{2} that the
dynamo term in Eq. (\ref{2.9}) has a simple electrodynamical meaning,
namely, it can be interpreted as a mean ohmic current directed along
the magnetic field:
\begin{equation}
\vec{J} = - \alpha \vec{H}.
\label{2.10}
\end{equation}
This equation tells us that an ensemble of screw-like vortices with
zero mean helicity is able to generate loops in the magnetic flux
tubes in a plane orthogonal to the one of the original field.
This observation will be of some related interest for the physical
interpretation of the results we are going to present in the following
paragraph. We finally notice that if the velocity field {\em is}
parity-invariant (i.e. no vorticity for scales comparable with the
correlation length of the magnetic field), then the dynamics of the
infrared modes is decoupled from the velocity field since, over those
scales, $\alpha =0$.

\subsection{AMHD equations for electroweak plasmas}

The electroweak plasma in {\em complete} thermal equilibrium at a temperature
$T$ can be characterized by $n_f$ chemical potentials
$\mu_i,~i=1,..., n_f$ corresponding to the exactly conserved global charges  
\begin{equation}
N_{i} = L_{i} - \frac{B}{n_{f}}
\label{2.11}
\end{equation}
($L_{i}$ is the lepton number of the $i$-th generation, $B$  is the baryon
number, and $n_{f}$ is the number of fermionic generations). One should
also introduce a chemical potential $\mu_Y$ corresponding to weak
hypercharge, but its value is fixed from the requirement of the
hypercharge neutrality of the plasma, $\langle Y \rangle =0$.

We want to
study this plasma slightly out of thermal equilibrium; in particular,
we want to see what happens with a non-uniform distribution of the
hypermagnetic field. Because of the  anomaly, and thanks to the 
 arguments illustrated  in Section 2.1,
 this field is coupled to the fermionic number densities. In
principle, different chemical potentials can be assigned to all the 
fermionic degrees of freedom of the electroweak theory ($45$ if
$n_{f}=3$) and the coupled system of Boltzmann-type 
equations for these chemical potentials and the hypercharge fields may be
written. Since we are interested in the slow processes in the
plasma, this is not necessary. If the coupling, corresponding to some
slow process, is switched off, then the  electroweak theory acquires an
extra conserved charge and  a further  chemical potential should be added
to the system  given in Eq.  (\ref{2.11}). 

An interesting observation (which turns out to be quite important in our
 context) has been made in
\cite{42}, where it was noticed that perturbative reactions with
right-handed electron chirality flip are out of thermal equilibrium at
temperatures higher than some temperature $T_R$.\footnote{ This
temperature depends on the particle physics model, see also the 
 discussion reported in Section 5. In the MSM $T_R \simeq 80$ TeV \cite{42}.} 
Thus, the number of right electrons  is
perturbatively conserved at temperatures $T>T_R$ 
and the  chemical potential $\mu_R$ can be introduced for
it. On the other hand, this charge  is not conserved 
because of the  Abelian anomaly,
\begin{equation}
\partial_\mu j^\mu_R = -\frac{g'^2 y_R^2}{64 \pi^2} {\cal Y}_{\mu\nu}
\tilde{{\cal Y}}^{\mu \nu},
\end{equation}
and it is therefore coupled  to the hypermagnetic field.
Here  ${\cal Y}$ and $\tilde{{\cal Y}}$ are, respectively, the $U_Y(1)$ 
hypercharge field strengths and their duals, $g'$ is the associated gauge
coupling and $y_R=-2$ is the hypercharge of the right electron. 

Now we are ready to derive the anomalous MHD equations \cite{14, us}.
The effective Lagrangian density describing
the dynamics of the gauge fields at finite fermionic density is \cite{26}:
\begin{equation}
{\cal L}_{Y, e_{R}} = -\frac{1}{4}\sqrt{-g}
Y_{\alpha\beta}Y^{\alpha\beta} - \sqrt{-g} J_{\alpha}Y^{\alpha} 
+ \mu \epsilon_{ijk} Y^{ij} Y^{k},~~~\mu = \frac{g'^2}{4\pi^2}\mu_{R}
\label{2.15}
\end{equation}
($g$ is the determinant of the metric defined in (\ref{2.1});
$Y_{\alpha\beta} = \nabla_{[\alpha}Y_{\beta]}$; 
$\nabla_{\alpha}$ is the covariant
derivative with respect to the metric (\ref{2.1})[notice that in the
metric (\ref{2.1}) 
$\nabla_{[\alpha}Y_{\beta]} = \partial_{[\alpha}Y_{\beta]}$]; $g'$ is the
Abelian coupling constant).
The first and the last terms in Eq. (\ref{2.15}) are nothing but the
curved space generalization of the flat-space effective Lagrangian for
the hypercharge fields at finite fermion density \cite{26},
$J_{\alpha}$ is the ohmic current.
The equations of motion for the
hyperelectric and hypermagnetic fields are then
\begin{eqnarray}
\frac{\partial{{\vec{H}}_{Y}}}{\partial\tau} = -\vec{\nabla}
\times {\vec{E}}_{Y}
,~~~~~
\frac{\partial{\vec{E}}_{Y}}{\partial\tau}+ {\vec{J}}_{Y} =
\frac{g'^2}{\pi^2} \mu_{R} a {\vec{H}}_{Y}+
{\vec{\nabla}}\times{ \vec{H}}_{Y}, 
\nonumber\\
{\vec{\nabla}}\cdot{\vec{H}}_{Y}=0,~~~~~~~~ ~~~~~~~~~~~~
{\vec{\nabla}}\cdot {\vec{E}}_{Y}=0,
\nonumber\\
{\vec{\nabla}}\cdot{\vec{J}}_{Y}=0,~~~ ~~~~
{\vec{J}}_{Y}=\sigma ({\vec{E}}_{Y} +
\vec{v}\times{\vec{H}}_{Y}),~~\sigma=\sigma_{c} a(\tau)
\label{2.16}
\end{eqnarray}
(${\vec{E}}_{Y} = a^2 {\vec{\cal E}}_{Y}$, 
${\vec{H}}_{Y} = a^2 {\vec{\cal H}}_{Y}$,
${\vec{{\cal H}}}_{Y}={\vec{\nabla}}\times {\vec{{\cal Y}}}$,
${\vec{{\cal E}}}_{Y} = - \frac{\partial}{\partial t}{\vec{{\cal
Y}}})$. For the EW theory the conductivity of plasma is 
$\sigma \simeq 70 T$ \cite{cond}.
To the  equations of motion of  the hypercharge field (\ref{2.16}) 
we have to add the evolution equation of the right electron chemical
potential, which accounts for the anomalous and perturbative 
non-conservation of the right electron number density ($n_R$) :
\begin{equation}
 \frac{\partial n_{R}}{\partial t} = - \frac{g'^2
}{4\pi^2} 
{\vec{{\cal E}}}_{Y}\cdot{\vec{{\cal H}}}_{Y} - \Gamma (n_{R}-n_R^{eq}),
\label{nr}
\end{equation}
where $\Gamma$ is the perturbative chirality-changing rate, 
$\Gamma = T\frac{T_R}{M_0}$, $n_R^{eq}$ is the equilibrium value of
the right electron number density,
and the term proportional to ${\vec{E}}_{Y}\cdot{\vec{H}}_{Y}$ is 
the  right electron anomaly contribution. 

Finally, the relationship between the right electron number density
and the chemical potential must  be specified. This relation depends
upon  the particle content of the theory, e.g. upon  the number of fermionic
generations, the number of Higgs doublets, etc. We will write it for
the case of the MSM \cite{42b}: the specific coefficients change only  
slightly for other theories and do not have a significant 
impact on the results. For generality, we assume that the Universe is
asymmetric not only with respect to the number of right electrons, but
also with respect to corresponding conserved charges defined in 
(\ref{2.11}) and compute all the relevant chemical potentials:
\begin{eqnarray}
\mu_{R} &=& \frac{2}{45} \pi^2 N_{eff} [ \frac{783}{88} \delta_{R} -
\frac{201}{88} \delta_{1} +\frac{15}{22} (\delta_{2} + \delta_{3})] T,
\nonumber\\
\mu_{1} &=&  \frac{2}{45} \pi^2 N_{eff} [ -\frac{201}{88} \delta_{R} +
\frac{1227}{440} \delta_{1} +\frac{3}{110} (\delta_{2}+ \delta_{3})] T,
\nonumber\\
\mu_{2} &=&  \frac{2}{45} \pi^2 N_{eff} [\frac{15}{22} \delta_{R} +
\frac{3}{110}\delta_{1} + \frac{14}{55} \delta_{2} +
\frac{124}{55}\delta_{3}] T,
\nonumber\\
\mu_{3} &=& \frac{2}{45} \pi^2 N_{eff} [ \frac{15}{22} \delta_{R} + 
\frac{3}{110} \delta_{1} + +\frac{14}{55} \delta_{2} +
\frac{124}{55}\delta_{3}]T,
\nonumber\\
\mu_{Y} &=& \frac{2}{45} \pi^2 N_{eff} [\frac{27}{88}\delta_{R} +
\frac{11}{440} \delta_{1} +\frac{39}{110} (\delta_{2} +\delta_{3})]T,
\label{2.14}
\end{eqnarray}
where $\delta_{i}$ is the asymmetry of $i$-th conserved charge,
$\delta_{R} = n_{R}/s$ ($s$ is the entropy density)
is the right electron asymmetry, $N_{eff} = 106.75$ is the effective
number of relativistic degrees of freedom in the symmetric phase of
the MSM.

With the use of relations (\ref{2.14}), Eq. (\ref{nr}) can be
rewritten completely in terms of the right electron chemical potential: 
 \begin{equation}
\frac{1}{a} \frac{\partial (\mu_{R} a)}{\partial \tau} = - \frac{g'^2
}{4\pi^2} \frac{783 }{88} \frac{ 1}{a^3 T^3}
{\vec{E}}_{Y}\cdot{\vec{H}}_{Y} - \Gamma (\mu_{R} a).
\label{mur}
\end{equation}

At finite hyperconductivity (in what
we would call, in a MHD context, ``resistive'' approximation)
we have that from Eq. (\ref{2.16}) the induced
hyperelectric field is not exactly orthogonal to the hypermagnetic one and,
moreover, an extra ``fermionic'' current comes in the game thanks to
the fact that we are working at finite chemical potential. 
Therefore in our context the resistive Ohm law can be written as
\begin{equation}
{\vec{E}}_{Y} = \frac{{\vec{J}}_{Y}}{\sigma}
-{\vec{v}}\times{ \vec{H}}_{Y}
 \simeq \frac{1}{\sigma}\left( 
\frac{4 \alpha'}{\pi} \mu_{R} a{\vec{H}}_{Y}+
{\vec{\nabla}}\times{ \vec{H}}_{Y}\right)
-{\vec{v}}\times{ \vec{H}}_{Y}, ~~~\alpha' =\frac{g'^2}{4\pi}~.
\label{ohm}
\end{equation}

In the bracket appearing in 
 Eq. (\ref{ohm}) we can identify two different contributions. One is
associated with the curl of the magnetic field. We will call this the MHD
contribution, since it appears in the same way in ordinary plasmas.
The other contribution contains the chemical potential and it is
directly proportional to the magnetic field and to the chemical
potential. This is a typical finite density effect. In fact the extra
Ohmic current simply describes the possibility that the energy sitting
in real fermionic degrees of freedom  can be transferred to the 
 hypermagnetic field. 
It may be of some interest to
notice the analogy between the first term of Eq. (\ref{ohm}) and the
typical form of the ohmic current discussed in Eq. (\ref{2.10})
appearing in the context of the dynamo mechanism. In the latter case the
presence of a current (proportional to the vorticity through the
$\alpha$ dynamo term)  was indicating that large length scales 
magnetic fields could grow by eating up fluid vortices. 
By inserting ${\vec{E}}_{Y}$ obtained from the generalized Ohm law
(\ref{ohm}) in the evolution equations (\ref{2.16}) of the
hypercharge fields, we obtain the generalized form of the magnetic 
diffusivity equation  (\ref{2.7}):
\begin{equation}
\frac{\partial{{\vec{H}}_{Y}}}{\partial \tau} =- \frac{4 a
\alpha'}{\pi\sigma}  
\vec{\nabla}\times\left({\mu_{R} \vec{H}}_{Y}\right) 
+{\vec{\nabla}}\times(\vec{v}\times{\vec{H}})
+ \frac{1}{\sigma}
 \nabla^2 {\vec{H}}_{Y}.
\label{hyperdiffusivity}
\end{equation}
In order to be consistent with our resistive approach
 we neglected terms
containing time derivatives of the electric field, which are 
sub-leading provided the conductivity is finite. In our considerations
 we will also make a further simplification, namely we will assume
 that the EW plasma is (globally) parity-invariant and that,
 therefore, no global vorticity is present. Therefore, since the
 length scale of variation of the bulk velocity field is much shorter
 than the correlation distance of the hypermagnetic field, the
 infrared modes of the hypercharge will be practically unaffected by
 the velocity of the plasma, which will be neglected when the large-scale part
 of the hypercharge is concerned.  This corresponds to the usual MHD
 treatment of a mirror symmetric plasma (see, e.g. Eq. (\ref{2.9}), 
when $\alpha =0$).

Eqs. (\ref{hyperdiffusivity}) and (\ref{mur}) form a set of
AMHD equations for the hypercharge magnetic field and right electron
chemical potential in the expanding Universe.

An important property of Eqs. (\ref{2.16}) and (\ref{hyperdiffusivity})
is that they are {\em not} conformally invariant. The 
conformal invariance of the ordinary Maxwell equations implies that 
the equations for the rescaled fields  in curved space 
keep the same form also in flat space  in terms of the non-rescaled
fields provided the conformal time coordinate is employed in curved space and
the cosmic time coordinate is employed in flat space. 
We can easily see that this is not the case of Eq. (\ref{2.16}) by
 writing our evolution equations 
in flat space:
\begin{eqnarray}
\frac{\partial{{\vec{{\cal H}}}_{Y}}}{\partial t} = -\vec{\nabla}
\times {\vec{{\cal E}}}_{Y}
,~~~~~
\frac{\partial{\vec{{\cal E}}}_{Y}}{\partial t}+
{\vec{j}}_{Y} = \frac{g'^2}{\pi^2}\mu_{R} 
{\vec{{\cal H}}}_{Y} +
{\vec{\nabla}}\times{ \vec{{\cal H}}}_{Y}, 
\nonumber\\
{\vec{\nabla}}\cdot{\vec{{\cal H}}}_{Y}=0,~~~~~~~~ ~~~~~~~~~~~~
{\vec{\nabla}}\cdot {\vec{{\cal E}}}_{Y}=0,
\nonumber\\
{\vec{\nabla}}\cdot{\vec{j}}_{Y}=0,~~~ ~~~~
{\vec{j}}_{Y}=\sigma_{c} {\vec{{\cal E}}}_{Y}. 
\label{2.17}
\end{eqnarray}
The lack of conformal invariance comes from the presence of the scale
factor in front of the right electron chemical potential 
in the evolution equation 
(\ref{2.16}) for ${\vec{ H}}_{Y}$.
Clearly, the explicit breaking of conformal invariance is also  reflected in
the Ohm law and in the hypermagnetic diffusivity equation which,
passing from curved to flat space, become
\begin{eqnarray}
{\vec{{\cal E}}}_{Y} &=& \frac{{\vec{j}}_{Y}}{\sigma_{c}}
\simeq \frac{1}{\sigma_{c}}\left( \frac{\alpha'}{\pi}\mu_{R} 
{\vec{{\cal H}}}_{Y} + {\vec{\nabla}}\times{ \vec{{\cal
H}}}_{Y}\right),
\label{flatohm}\\
\frac{\partial{{\vec{{\cal H}}}_{Y}}}{\partial t} &=& 
-\frac{4 \alpha'}{\pi\sigma}  
\vec{\nabla}\times\left({\mu_{R} \vec{{\cal H}}}_{Y}\right) 
+ \frac{1}{\sigma_{c}} \nabla^2 {\vec{{\cal H}}}_{Y}.
\label{flatdiffusivity}
\end{eqnarray}
In flat space the kinetic equation of the right electron chemical
potential becomes instead:
\begin{equation}
\frac{\partial \mu_{R}}{\partial t} =
 - \frac{g'^2}{4~\pi^2~T^2} \frac{783 }{88} 
{\vec{{\cal E}}}_{Y}\cdot{\vec{{\cal H}}}_{Y} - \Gamma \mu_{R}.
\label{2.23}
\end{equation}
It is interesting to notice that the term
containing the chemical potential in Eq. 
(\ref{hyperdiffusivity}) plays a role similar to that of the dynamo
term, since it also produces an instability \cite{14}. Its physical
interpretation  is actually quite simple. We could
define, as in the case of the ordinary MHD, a generalized Reynolds
number that measures the relative weight of the two terms on the
r.h.s. of Eq. (\ref{hyperdiffusivity}). If the diffusion term
 dominates, then the
flux of magnetic hypercharge will diffuse through the plasma. If on
the contrary we are in the inertial range (where the 
diffusivity term is negligible) there are two possibilities. If the
chemical potential is exactly zero then the hypermagnetic field will be
frozen into the plasma element as required by magnetic flux
conservation. However if we
are at finite fermion density 
the energy density  sitting in fermionic
degrees of freedom may be  transformed in infrared modes of the
hypermagnetic field.

\section{Fermions from the hypercharge field}

\subsection{$T>T_c$}

In this section we are going to compute the relationship between
hypermagnetic fields and  induced fermionic chemical potential,
at temperatures larger than the critical temperature of the
electroweak phase transition $T_c$. 

There is an important consequence of the resistive approximation. 
By using the Ohm law given by
Eq. (\ref{flatohm}) we can eliminate the 
hyperelectric field from the kinetic equation of the right electrons
and obtain 
\begin{equation}
\frac{\partial}{\partial t}\left(\frac{\mu_R}{T}\right)=
-\frac{g'^2}{4\pi^2 \sigma_{c}T^3} \frac{783}{88}
{\vec{{\cal H}}}_{Y}\cdot \vec{\nabla}\times
{\vec{{\cal H}}}_{Y} - (\Gamma + \Gamma_{{\cal H}} ) \frac{\mu_R}{T},
\label{flatkinetic}
\end{equation}
where
\begin{equation}
\Gamma_{{\cal H}} = \frac{783}{22} \frac{\alpha'^2}{\sigma_{c}\pi^2}
\frac{|{\vec{{\cal H}}}_{Y}|^2}{T^2}.
\label{2.24}
\end{equation}
We notice immediately that the source term appearing in the r.h.s. of Eq.
(\ref{flatkinetic}) (and coming from the anomaly) is indeed strongly
reminiscent of what we would call (in a MHD context) magnetic
helicity.
From Eq. (\ref{flatkinetic}) one can see that the right-electron number is not
conserved (even if $\Gamma=0$) because of the Abelian anomaly, provided a
non-zero hypermagnetic field is present (cf. Ref. \cite{krs}).
Eq. (\ref{flatkinetic}) can be solved in the  adiabatic
approximation at temperatures $T < T_R$, when perturbative 
right electron chirality flip reactions are in thermal
equilibrium. Neglecting the time derivative of the chemical potential,
we get:
\begin{equation}
\frac{\mu_R}{T} \simeq -\frac{\alpha'}{\pi\sigma_{c}T^3}\frac{783}{88}
\frac{{\vec{{\cal H}}}_{Y}\cdot \vec{\nabla}\times
{\vec{{\cal H}}}_{Y} }{\Gamma +\Gamma_{{\cal H}}}.
\label{2.25}
\end{equation}
The solution
(\ref{2.25}) can be inserted into Eq. (\ref{flatdiffusivity}) 
for the magnetic
field, which will become a partial (non-linear) differential
 equation containing only the hypermagnetic field. 
Thus, an inhomogeneous hypermagnetic field
can produce a spatial variation in the chemical content of the plasma.
In fact, according to Eq. (\ref{2.14}), spatial fluctuations in the
right electron chemical potential will determine fluctuations not only
in the right electron number density but also in the number densities
associated with the other fermion asymmetries. 
Fluctuations in the number density of some species are 
frequently  called isocurvature 
perturbations. There are actually two
regimes where Eq. (\ref{2.25}) can be analysed. The first one is the
regime where $\Gamma > \Gamma_{{\cal H}}$. In this case the rate of
right electron dilution is essentially determined by the perturbative
processes, which can flip the chirality of the right electrons.
In the opposite case ($\Gamma_{{\cal H}}> \Gamma$) the rate is mainly
due to the presence of the Abelian anomaly, and 
\begin{equation}
n_{R} \simeq -\frac{88 \pi^2}{783 g'^2}\left( \frac{{\overline{{\cal
H}}}_{Y}\cdot\overline{\nabla}\times {\overline{{\cal H}}}_{Y} }
{ |{\overline{{\cal H}}}_{Y}|^2  }\right) +
O\left(\frac{\Gamma}{\Gamma_{{\cal H}}}\right),
~~~\Gamma_{{\cal H}}> \Gamma ~~. 
\label{2.26}
\end{equation}
In Eq. (\ref{2.26})  the chirality-changing  rate 
only comes in the correction. Moreover, since the
hypermagnetic  field intensity appears with the same power
in the numerator and in the denominator, 
the right electron fluctuations are  independent
of the magnitude of the hypermagnetic field  fluctuations and 
fixed by their spatial distribution.
It is interesting to notice that in Eq. (\ref{2.26}) the actual value of
the conductivity completely cancels and only appears in the
correction.

\subsection{$T<T_c$}

Now we are going to discuss what happens after the electroweak phase 
transition. Below $T_c$ the hypercharge magnetic
fields are converted into ordinary magnetic fields. The latter
 do not have anomalous coupling to fermions, and the usual
MHD equations are fully valid. Any source term that was inducing a
non-vanishing chemical potential disappears. Thus, the
transformation of the magnetic fields into fermions is no longer
possible. It seems, therefore, that the matter fluctuations after the
phase transition will be given by the fluctuations right before the
phase transition. The last statement is in fact wrong for two
reasons. First, if the phase transition is weakly first order, so that
sphaleron processes are in thermal equilibrium after it, then any
fluctuations of the fermionic charges will disappear. In this
particular case all anomalous effects that existed before $T_c$ are simply
``forgotten'', since the system passes through an equilibrium period with
respect to fermion number non-conservation. 
Let us admit 
that the electroweak phase transition is strongly first order and a
necessary condition for EW baryogenesis \cite{25} is satisfied. Then, 
there is an important  ``storage" effect (and we come to the second
point), which amplifies the
estimates of Eqs. (\ref{2.25}) and (\ref{2.26}) by many orders of
magnitude. The point is
that the fermion number can sit not only in the fermions 
(and in their associated chemical potential), 
but also in the hypermagnetic field itself. 
At the EW phase transition, this fermion number must be released in the
form of real fermions, just because the ordinary magnetic field, which
survives after transition, cannot carry fermion number. To compute the
density of the Chern-Simons number $n_{CS}$ of the 
hypercharge field configuration before the
EW phase transition we just integrate
${\vec{\cal E}}_{Y}\cdot{\vec{\cal H}}_{Y}$ over the  time:
\begin{equation}
\Delta n_{CS}(t_{c}) = -\frac{y^2_{R}~g'^2}{16 \pi^2} 
\int_{0}^{t_{c}}{\vec{\cal E}}_{Y}\cdot{\vec{\cal H}}_{Y}d t.
\label{5}
\end{equation}
In order to estimate  this integral we have to solve the coupled
system given by Eqs. (\ref{flatdiffusivity}) and
(\ref{flatkinetic}). The main contribution to this integral comes from
the largest time $t \sim t_{c}$, where reactions with right electron
chirality flip are in thermal equilibrium. Thus,  we can
use again the adiabatic approximation (which implies that
$\frac{\partial {\mu}_{R}}{\partial t} \sim 0$) and obtain:
\begin{eqnarray}
\Delta n_{CS}(t_{c}) &=&-\frac{\alpha'}{2\pi}
\int_{0}^{t_{c}} \frac{\Lambda(\vec{x},t)}{\Gamma +
\Gamma_{\cal H}}\frac{\Gamma}{\sigma_{c}}~dt,
\nonumber\\
\Lambda(\vec{x},t) &=& {\vec{{\cal H}}}_{Y}\cdot \vec{\nabla}\times
{\vec{{\cal H}}}_{Y}.
\label{integral}
\end{eqnarray}

This Chern-Simons number will be released at the EW phase transition
in the form of fermions, which will not be destroyed by the sphalerons
if the phase transition is strongly first order. The density of the
baryonic number $n_B$ is just given by integrated anomaly:
\begin{equation}
n_{B}(t_{c}) = - \frac{n_{f}}{2} \Delta n_{CS}(t_{c}).
\label{baryon}
\end{equation}
This equation is our main result.
Once the  hypermagnetic background is specified, the time integration
appearing in (\ref{integral}) can be performed. If the typical scale of the
configuration is larger than the magnetic diffusivity distance
\begin{eqnarray}
r_{\sigma} \sim \frac{1}{k_{\sigma}}\sim 10^{-9}~\times L_{ew},
\label{diffusivityscale}
\end{eqnarray}
where $L_{ew}\sim 3~{\rm cm}$ is the size of the EW horizon at
$T_{c}\sim~100~{\rm GeV}$, then all the modes of the hypermagnetic
field with momentum $k$ smaller than 
\begin{eqnarray}
k_{\sigma} \sim \sqrt{\frac{\sigma_{c}}{M_{0}}} T,~~~M_{0} =
\frac{M_{Pl}}{1.66~ \sqrt{N_{eff}}} \sim 7.1~\times ~10^{17}~{\rm GeV}
\nonumber
\end{eqnarray}
will remain frozen into  the EW  plasma element.
Thus the baryon-number fluctuations 
can be written as 
\begin{equation}
\delta\left(\frac{n_{B}}{s}\right)(\vec{x}, t_{c}) =
\frac{\alpha'}{2\pi\sigma_c}\frac{n_f}{s}
\frac{{\vec{{\cal H}}}_{Y}\cdot \vec{\nabla}\times
{\vec{{\cal H}}}_{Y}}{\Gamma +\Gamma_{{\cal H}}}\frac{\Gamma M_0}{T_c^2},~~
\Gamma_{{\cal H}} = \frac{783}{22} \frac{\alpha'^2}{\sigma_{c}\pi^2}
\frac{|{\vec{{\cal H}}}_{Y}|^2}{T_{c}^2}.
\label{final}
\end{equation}
Notice that in Eq. (\ref{final}) there is an enhancement by 
a factor $ \sim \Gamma M_0/T_c^2$ arising from the
time integration of the anomaly term. We also point out that  
for $\Gamma_{\cal H} \laq \Gamma$ the rate of right electron 
chirality flip cancels out.
This last expression can be easily written in terms of the 
corresponding curved
space quantities and the only point to be kept in mind is that,
in curved space, the chemical potential is multiplied by the scale
factor (which breaks the conformal invariance of the AMHD equations).

\renewcommand{\theequation}{4.\arabic{equation}}
\setcounter{equation}{0}
\section{Stochastic hypermagnetic backgrounds}
Two qualitatively different classes of
 hypermagnetic  backgrounds can be studied. 
The first class is characterized by a
non-vanishing magnetic helicity (i.e.  
$\langle \vec{\cal H}_{Y} \cdot \vec{\nabla}\times \vec{\cal
H}_{Y}\rangle \neq 0$),
which implies that the hypercharge field is topologically
non-trivial and parity-non-invariant. Therefore, in this class of
backgrounds not only fluctuations in
the baryon number will be produced, but also the generation of the
baryon asymmetry is possible. We will
discuss this possibility later in Section 6. 

The aim of this section is to relate the properties 
of stochastic background with
zero average magnetic helicity to the baryon number fluctuations. For
this type of magnetic field  
$\langle \delta(n_{B}/s)(\vec{x},t_{c})\rangle=0 $ but 
$\langle
\delta(n_{B}/s)(\vec{x}+\vec{r},t_{c})\delta(n_{B}/s)(\vec{x},t_{c}
)\rangle \neq 0$, 
so that only the inhomogeneities of baryonic number are produced.
We will be interested here in the formal aspect of this relation,  and
we will focus our attention on the study of the phenomenological 
applications in Section 5.

Consider a stochastic hypermagnetic field whose (parity-invariant)
two-point function is,
\begin{equation}
G_{ij}(\vec{r}) = \langle H_{i}(\vec{x}) H_{j}(\vec{x}+\vec{r})\rangle
= \int e^{i \vec{k}\cdot\vec{r}} G_{ij}(k) d^{3}k,
\label{4.1}
\end{equation}
where, because of  transversality of the magnetic field 
\begin{equation}
G_{ij}(k) = k^2 f(k) \left(\delta_{ij} - \frac{k_{i} k_{j}}{k^2}\right)~.
\label{4.3}
\end{equation}
The average appearing in (\ref{4.1}) denotes an ensemble average.
As was previously stated,  the average
 hypermagnetic helicity is $\langle\Lambda(\vec{x},t_{c})\rangle
=\langle \vec{H}\cdot\vec{\nabla}\times\vec{H} \rangle =0$
in the case of the transverse and parity-invariant 
two-point function given in Eqs. (\ref{4.1}) and (\ref{4.3}). 

The
assumption of the stochasticity of the background implies that the
higher-order correlation functions of the magnetic hypercharge fields
can be computed in terms of the two-point function (\ref{4.1}).
For example,  the four-point function
can be completely expressed in terms of the two-point function (\ref{4.1}):
\begin{eqnarray}
& &\langle H_{k}(\vec{x}')
H_{j}(\vec{x}) H_{l}(\vec{y}') H_{n}(\vec{y}) \rangle=
 [ \langle H_{k}(\vec{x}')H_{j}(\vec{x}) \rangle
\langle H_{l}(\vec{y}') H_{n}(\vec{y}) \rangle +
\nonumber\\
& & \langle H_{k}(\vec{x}') H_{l}(\vec{y}')\rangle
\langle H_{j}(\vec{x}) H_{n}(\vec{y})\rangle +
\langle H_{k}(\vec{x}') H_{n}(\vec{y})\rangle\langle
H_{j}(\vec{x}) H_{l}(\vec{y}') \rangle~].
\label{4points}
\end{eqnarray}

We are now going to  compute the level of the induced fluctuations
by the above-mentioned stochastic
hypermagnetic background.
We parametrize  the spectral properties of our correlation
function by assuming a power law behaviour of its Fourier transform:   
\begin{equation}
f(k)=\frac{1}{k}\left(\frac{k}{k_{1}}\right)^{\alpha}\exp{ \Biggl[-2 
\left(\frac{k}{k_{\sigma}}\right)^2\Biggr]}.
\label{4.4}
\end{equation}
This representation only depends upon two unknown parameters, namely the
slope ($\alpha$) and the amplitude, which can be changed by changing
 $k_{1}$. The exponential damping appearing in the mode
function is not the result of any assumption, but it is a direct
consequence of the fact that, according to the hyperdiffusivity
equations (\ref{hyperdiffusivity}) and (\ref{flatdiffusivity}), all the
modes $k> k_{\sigma}$ decay  thanks to the finite 
value of the conductivity.
The hypermagnetic energy density is obtained by tracing the Green
function defined in Eq. (\ref{4.1}) for $\vec{r}=0$,
\begin{equation}
\langle|\vec{H}(\vec{x})|^2\rangle = {\rm Tr}[G_{ij}(0)]= 2 \int d^3 k k
\left(\frac{k}{k_{1}}\right)^{\alpha} 
\exp{\Biggl[-2\left(\frac{k}{k_{\sigma}}\right)^2\Biggr]}.
\label{4.5}
\end{equation}
Because of the exponential damping, this integral
is always ultraviolet convergent and can be very simply performed:
\begin{equation}
\langle \frac{|\vec{H}(\vec{x})|^2}{T^4}\rangle = 4\pi\xi^{-\alpha} 
2^{-\frac{\alpha +4}{2}}\Gamma\left(\frac{\alpha + 4}{2}\right) 
\left(\frac{k_{\sigma}}{T}\right)^{\alpha + 4},
\label{energy}
\end{equation}
where $\xi = k_{1}/T$.

We will often have  to compute various four-point functions, and it is
sometimes of great help to evaluate the higher-order Green functions
not in Fourier space (where complicated convolutions would  appear) but
directly in real space. A generic rotationally and parity invariant
Green  function can always be written in real space as  
\begin{equation}
G_{ij}(|\vec{r}|) = F_{1}(r)\delta_{ij} + r_{i}r_{j} F_{2}(r),~|\vec{r}|=r,
\label{realspace}
\end{equation}
where 
\begin{equation}
F_{1}(r) = \frac{\partial}{\partial r^2} [ r^2 h(r^2)],~~~
F_{2}(r) = -\frac{\partial}{\partial r^2} [h(r^2)],~~~
r {\cal G}(r)=\frac{\partial}{\partial~r^2}[r^3~h(r^2)],
\label{F1F2}
\end{equation}
and ${\cal G}(r)$ is nothing but the trace of our two-point function, namely
\begin{equation}
{\cal G}(r) = {\rm Tr}[G_{ij}(r)] = \frac{4\pi}{R} k_{\sigma}^4
\left(\frac{k_{\sigma}}{k_{1}}\right)^{\alpha} \int_{0}^{\infty} q^{(2
+ \alpha)} e^{-2 q^2} \sin{R q} dq
\label{4.6}
\end{equation}
(with $q = \frac{k}{k_{\sigma}}$ and $R= \frac{r}{r_{\sigma}}$).
Clearly, this representation is transverse (i.e. 
$\frac{\partial}{\partial r^{i}} G_{ij}(r)=0$),
and moreover the integral over the spectrum appearing in
Eq. (\ref{4.6}) can be exactly performed in terms of known special
functions:
\begin{equation}
{\cal G}(R) = {\cal G}(0) F\left(-\frac{1}{2}
-\frac{\alpha}{2}, \frac{3}{2},
\frac{R^2}{8}\right)\exp{\Biggl[-\frac{R^2}{8}\Biggr]},~~~{\cal G}(0)=
 4 \pi k_{\sigma}^4 \left(\frac{k}{k_{\sigma}}\right)^{\alpha}
2^{-3 -\frac{\alpha}{2}} \Gamma\left(2 + \frac{\alpha}{2}\right)
\label{4.8}
\end{equation}
($F(a,b,z)$ is the confluent hypergeometric function and $\Gamma(z)$
is the Euler gamma function \cite{57,58}).

Some physical considerations are now in order.
In our problem  the relevant scales are those are not erased
by the plasma conductivity, namely, 
from Eq. (\ref{diffusivityscale}), all the scales $r>r_{\sigma}$. 
Therefore, the physical limit of all our 
correlation functions will always be the large-$R$ limit.
Moreover, a physically realistic situation does correspond, in our
considerations, to the case where the Green functions are decaying
at large length scales. If the Green functions decay at large distances
we automatically exclude the possibility that the energy  spectrum
of the hypermagnetic inhomogeneities will have some peak at large
wave-length. 
The large scales (i.e. $R>1$) limit of the normalized trace of 
our Green functions  will then be given by 
\begin{equation}
 g(R) =
\frac{\Gamma(\frac{3}{2})}{\Gamma(-\frac{1 +\alpha}{2})} 2^{6 +
\frac{3}{2}\alpha} R^{-(\alpha + 4)},~~~ g(R)= \frac{{\cal
G}(R)}{{\cal G}(0)}.
\label{4.10}
\end{equation}
In $k$-space
the magnetic energy density per logarithmic interval of frequency
is defined as \cite{17,19} as $\rho(k) = d \rho_{H}/d \ln{k}$
(where $\rho_{H} = \frac{1}{2} \langle|\vec{H}(\vec{x})|^2\rangle$). 
Therefore in our case
$\rho(k) \sim k^4 (k/k_{1})^{\alpha}$ which implies that
 ``blue''( $\alpha \gaq -4$) or ``violet''
($\alpha \gg -4 $) logarithmic energy spectra correspond to the
physically interesting case of two-points functions decaying at large
scales whereas for $\alpha<-4 $ we have  ``red''
logarithmic energy  spectra which are connected with Green's 
functions decreasing at small scales.
 The case of
flat  logarithmic energy spectra ($\alpha\simeq  -4$)
may quite naturally appear in string cosmological models
\cite{19}.

It is important to point out that if the Green functions decay at
large distances, then $ g(R)<1$. This observation will be of
some relevance when we will have  to explicitly evaluate our
fluctuations, since $g(R)$ will turn out to be a useful and
natural expansion parameter (see Appendix B for details).

We can also give the large $R$ expression of $h(r^2)$ since it can easily  be
deduced according to Eq. (\ref{F1F2}): 
\begin{equation}
h(R) = {\cal G}(0) \frac{\Gamma(\frac{3}{2})}{\Gamma(\frac{3-\epsilon}{2})} 
\frac{2^{\frac{3}{2}\epsilon+1}}{3-\epsilon} R^{-\epsilon},
\label{4.23}
\end{equation}
where we defined $\alpha= - 4 + \epsilon $.

We are now ready to compute the level of baryon-number 
fluctuations induced by our
stochastic background of magnetic hypercharge fields, which is defined
by the correlation function
\begin{equation}
\Delta(r,\tau_{c}) = \sqrt{|\langle\delta
\left(\frac{n_{B}}{s}\right)(\overline{x}, t_{c})
\delta\left(\frac{n_{B}}{s}\right)(\overline{x}+\overline{r},
t_{c})\rangle|} .
\label{4.29}
\end{equation}
There are two
regimes where this calculation can be performed depending upon the
relative weight of the two rates appearing in Eq. (\ref{flatkinetic}).
If $\Gamma~\gaq~\Gamma_{{\cal H}}$ the major technical problem we have
to face is to evaluate the correlation function of the magnetic
helicity at two different points; this involves, ultimately, the
calculation of a four-point function. The algebraic details  of this
long but straightforward calculation are given in Appendix A. The
result is given in terms of the functions appearing in the real space
parametrization of our Green's functions given in Eq. (\ref{F1F2}):
\begin{eqnarray}
& & \biggl\langle\left(\vec{H}\cdot\vec{\nabla}
\times\vec{H}
\right)(\vec{x})
\left(\vec{H}\cdot\vec{\nabla}\times\vec{H}\right)
(\vec{x} +\vec{r}) \biggr\rangle = 
\nonumber\\
& & -\frac{4}{r}F_{1}(r) \frac{d F_{2}(r)}{dr} 
- 2 F_{1}(r) \frac{d^2 F_{1}(r)}{dr^2} 
+ 4 r^2 [F_{2}(r)]^2 + 2 r F_{1}(r) \frac{dF_{2}(r)}{d r}
\nonumber\\
& & - 6 r F_{2}(r) \frac{dF_{1}(r)}{dr} + 6 F_{1}(r) F_{2}(r) 
+ 2 \left(\frac{dF_{1}(r)}{dr} \right)^2.
\label{4.19}
\end{eqnarray}
In this and  the following formulae  we will often use the notation
${\vec{B}}(\tau_{c}) = {\vec{H}}(\tau_{c})/ T^2(\tau_{c})$, which is
convenient since in  $\vec{B}$ the time dependence of the scale
factors cancels and the only time dependence left is due to the
evolution of the effective number of the relativistic degrees of
freedom in the plasma, $N_{eff}(\tau)$.
Inserting now Eq. (\ref{4.23}) into Eq. (\ref{F1F2}) we obtain:
\begin{eqnarray}
&&\Delta(r,t_{c})= \frac{45 n_{f} \alpha'}{\pi^2 N_{eff}(\tau_{c})
}\frac{T_{c}}{\sigma_{c}}\frac{M_{0}}{T_{c}}\frac{\xi^{4-\epsilon}C(\epsilon)}
{(rT_{c})^{1+\epsilon}}(1 + O(\lambda)),~~~\xi = \frac{k_{1}}{T}
\nonumber\\
&&C(\epsilon)=\frac{2^{\epsilon-\frac{3}{2}}\Gamma(\frac{\epsilon}{2}) }
{\Gamma(\frac{3-\epsilon}{2})}~\sqrt{\frac{\pi\epsilon(\epsilon+2)}
{(3-\epsilon)}},~~~
\lambda\sim\left(\frac{\Gamma}{\Gamma_{\cal H}}\right)^2 
(\frac{r}{r_{\sigma}})^{-2\epsilon}.
\label{Delta}
\end{eqnarray}
Notice that this expression holds for large rate
(i.e. $\Gamma>\Gamma_{{\cal H}}$) and for large scales
(i.e. $g(R)<1$, $R>1$).

The second regime in which one may wish to compute the level of
induced fluctuations is the one where $\Gamma ~\laq~ \Gamma_{{\cal
H}}$. The  main mathematical problem will be to
evaluate the correlation function of the hypermagnetic helicity
divided by the hypermagnetic energy density.
This is of course a strongly non-linear object which we cannot 
compute exactly.
By working at large scales and in the hypothesis that the Green
functions are decaying at large distances, the bottom line of this
calculation (reported  in Appendix B) is :
\begin{eqnarray}
\biggl\langle\left(\frac{\vec{B}\cdot\vec{\nabla}\times\vec{B}}{|\vec{B}|^2}
\right)(\overline{x})
\left(\frac{\overline{B}\cdot\overline{\nabla}\times\overline{B}}
{|\vec{B}|^2}\right)
(\vec{x} +\vec{r}) \biggr\rangle \simeq 
\nonumber\\
\frac{\langle(\vec{B}\cdot\vec{\nabla}\times\vec{B})
(\vec{x})(\vec{B}\cdot\vec{\nabla}\times\vec{B})
(\vec{x}+\vec{r})\rangle}{ \langle |\vec{B}(\vec{x})|^2
\rangle^2} + O(g(R)).
\label{4.17}
\end{eqnarray}
Notice that to estimate the numerator appearing at the r.h.s. of
Eq. (\ref{4.17}), Eq. (\ref{4.19}) can be used together with the
considerations reported in Appendix A.
\renewcommand{\theequation}{5.\arabic{equation}}
\setcounter{equation}{0}
\section{Phenomenological implications}

Having set all the formalism for computing baryon-number fluctuations,
we now come to the physical consequences. As we argued in Section
3.2, these fluctuations survive after the electroweak phase
transition only if it is strongly first-order; we will assume that
this is the case. We will argue in Section 6.1 that strong enough
magnetic fields make an EW phase transition strongly first-order even
in the case of the minimal standard model. Otherwise, some extension
of it can be considered.

An essential quantity entering all expressions for baryon-number
fluctuations is the ratio between the perturbative and
non-perturbative rate of the right electron chirality
flip. Fluctuations are larger for $\Gamma > \Gamma_{\cal H}$. In
Sections 5.1 and 5.2 we will assume that this is indeed the case,
and  analyse this assumption in detail in Section 5.3. 

As a preliminary warm up let us estimate the amplitude of
 baryon-number fluctuations at the magnetic diffusivity scale for 
a flat spectrum of magnetic fields ($\epsilon \ll 1$). If the 
energy sitting in
the background magnetic field is comparable with the energy density of
the photons, $\langle {\vec{\cal H}}_{Y}^2 \rangle \sim T_{c}^4$, then for
the smallest possible scale $r \sim 1/k_\sigma \sim 10^{-9}\times$(EW
horizon $\simeq 3$ cm) we get, from Eq. (\ref{Delta}):
\begin{equation}
\Delta(r_{\sigma}, t_{c}) \sim \frac{\alpha'}{N_{eff}} 
\sqrt{\frac{M_{0}}{\sigma_{c}}}\sim 10^{3}.
\label{estimate}
\end{equation}
This estimate is certainly quite large and it is unlikely to be
correct, since for such
huge fluctuation the back-reaction of the created fermions on the
magnetic fields and on the dynamics of the electroweak phase
transition (which we ignored) must be taken into account. 
Nevertheless, it shows that
considerable inhomogeneities in the baryonic number are possible on
small scales. The estimate (\ref{estimate})  considerably
exceeds the measure of the baryon asymmetry of the
Universe $n_B/s \sim 10^{-10}$, thus small size matter--antimatter
 are possible at the EW scale. At the same time, for even larger
scales (possibly relevant for structure formation), the fluctuations
of Eq. (\ref{Delta}) are
quite minute (since their amplitude decreases with the distance as
$1/r^{1+\epsilon}$) and may be safely neglected. 

\begin{figure}
\begin{center}
\begin{tabular}{|c|c|}
      \hline
      \hbox{\epsfxsize = 7.5 cm  \epsffile{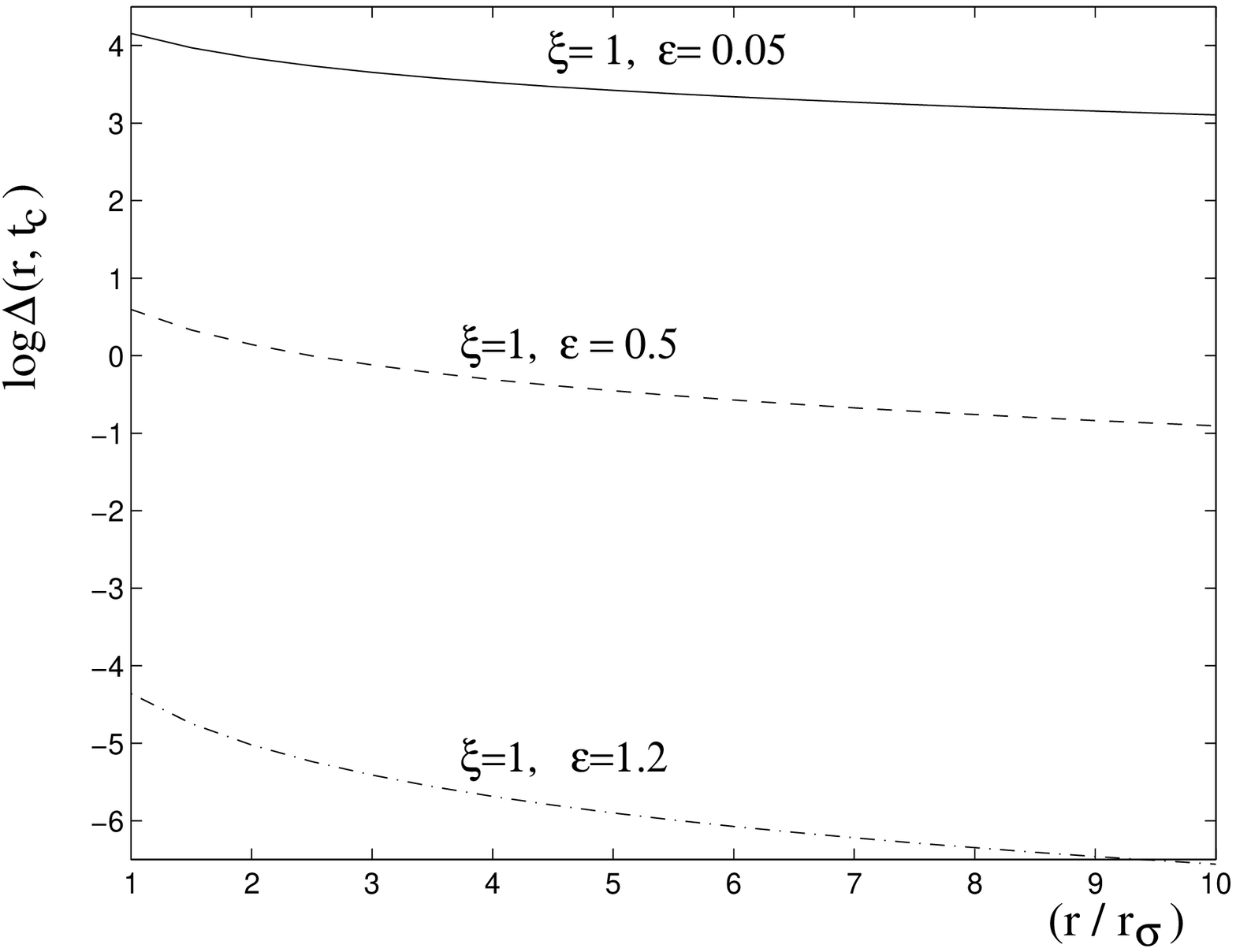}} &
      \hbox{\epsfxsize = 7.5 cm  \epsffile{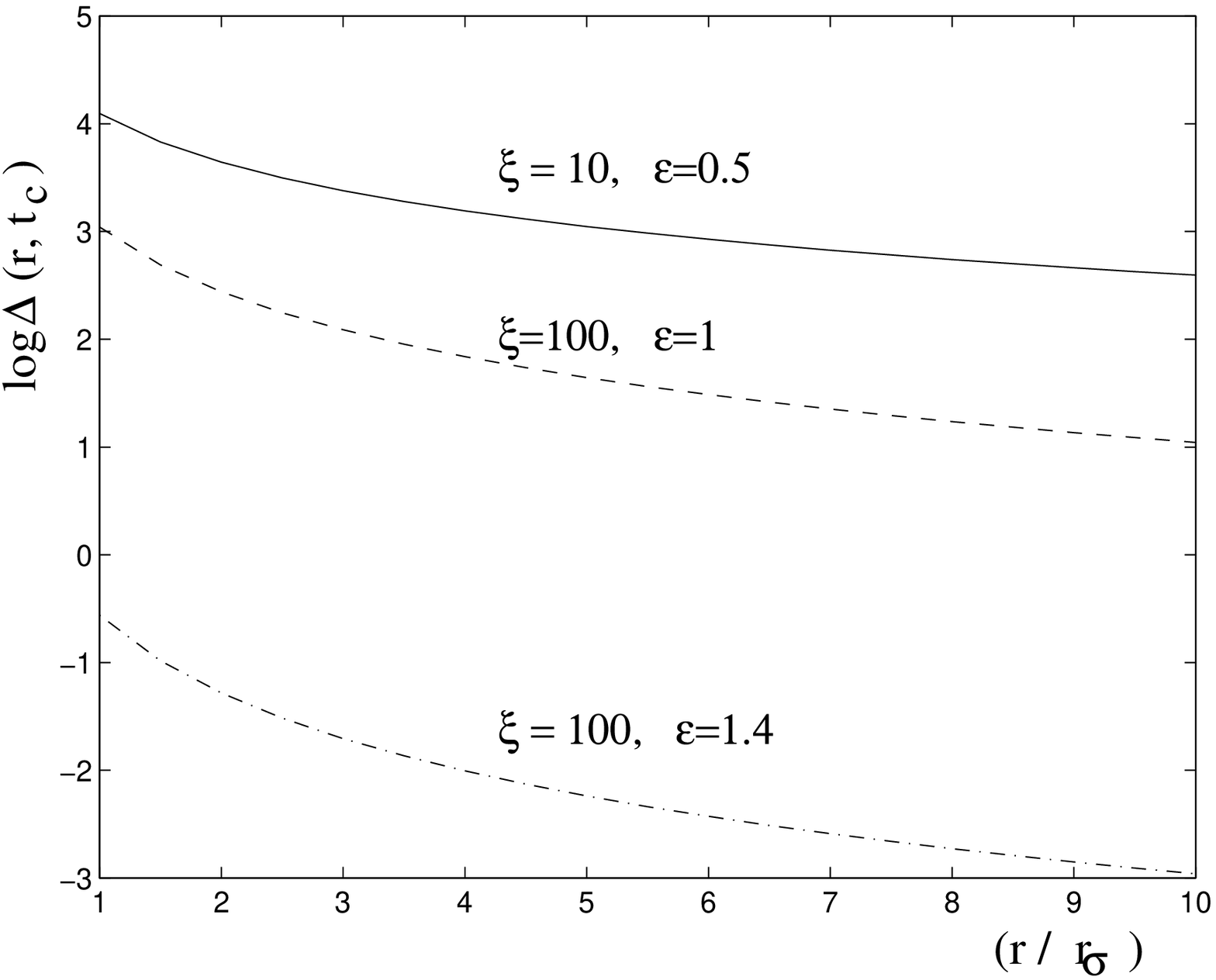}}\\
      \hline
\end{tabular}
\end{center}
\caption[a]{We plot the level of fluctuations given in Eq. (\ref{Delta})
for different values of amplitudes ($\xi$) and slopes ($\epsilon$) of
the hypermagnetic energy spectrum for scales slightly larger than the
diffusivity scale, which is $10^{-9} \times ~L_{EW}$ ($L_{EW}$ is the
EW horizon, see also Eq. (\ref{diffusivityscale})). These scales will
not be washed out by the finite value of the conductivity, and we can
immediately see that for a sufficiently blue spectrum ($\epsilon \ll
1$) $\Delta \sim 10^3$--$10^4$ as suggested by the estimate of Eq.
 (\ref{estimate}). Looking at the plots from top to bottom, we see
that  $\Delta$ decreases
 for increasingly violet spectra ($\epsilon~ \gaq ~1$) and for fixed
amplitude ($\xi$) of the hypermagnetic energy spectrum. We notice
that the results illustrated in this plot depend very weakly upon the
value of the hyperconductivity. In particular, in the present plots, we
assumed $\frac{\sigma_c}{T_c}\simeq 70 $ as fiducial value for the
hyperconductivity at the EW scale.}
\label{LogDelta}
\end{figure}

The above estimate
suggests that a quite natural outcome of the presence of stochastic
background of the primordial hypercharge field may be a rather
inhomogeneous distribution of matter and antimatter domains for scales
inside the EW horizon.
The fluctuations estimated in Eq. (\ref{Delta}) are also  illustrated in 
 {\bf Fig. \ref{LogDelta}} where the level of fluctuation is plotted
for different slopes and amplitudes of the hypermagnetic energy
spectrum. We clearly see  that by tuning $\xi$ (i.e. by tuning the
amplitude of the hypermagnetic energy spectrum) the level of induced
matter--antimatter fluctuations can be as large as suggested by the
estimate of Eq. (\ref{estimate}).

We will now  discuss  the  relevance of the 
 generated fluctuations for the BBN. In fact sizable
 matter--antimatter 
fluctuations can provide a
new type of initial conditions for non-homogeneous BBN. From a more
conservative point of view, we can instead assume that the BBN was
essentially homogeneous; then our considerations provide a
new bound on primordial magnetic fields present at the EW epoch. 
For completeness we will compare the bounds arising from the
occurrence of matter--antimatter domains for scales of the order of the
neutron diffusion distance with the bounds for magnetic fields that
are at present coherent over much larger scales. 

\subsection{BBN and matter--antimatter fluctuations}

The success of the homogeneous and isotropic
nucleosynthesis may impose strong constraints upon the baryon-number 
fluctuations possibly produced prior to the formation of the
light nuclei. 
Broadly speaking, the predictions of
homogeneous BBN for the 
primordial abundances of the light elements are compatible with the
observations only if the baryon-to-photon ratio lies in a quite narrow range
around  ${n}_B/n_\gamma = 3\times 10^{-10}~-~ 10^{-9}$ \cite{32}.

Generally, if $\Delta(r,t_{c}) > \frac{n_{B}}{s}$
for some length scale $r$, we have to conclude that
matter--antimatter domains will be formed. 
If, on the other hand,  $\Delta(r,t_{c}) < \frac{n_{B}}{s}$ at all scales,
only positive-definite
fluctuations in baryonic density  are produced. 

Upper and lower limits on the  scales over  which a  perturbation in baryon
number can affect nucleosynthesis  through neutron-proton segregation
are determined by the comoving diffusion lengths of neutrons and
protons at the beginning of nucleosynthesis.
At high temperatures the diffusion lengths of neutrons and protons are
almost the same, since neutron-proton equilibrium is guaranteed by
weak interactions. After weak interactions have fallen out of
equilibrium, nucleons retain their identity as neutrons and protons, and
diffusive segregation can occur.
Coulomb scattering between protons and electrons (or positrons) give a
cross section roughly equal to the Thompson cross section.
Since neutrons have a magnetic moment they scatter electrons with a
cross section of $8\times 10^{-31}~{\rm cm}^{2}$.
Neutrons scatter also nucleons and the scattering cross section 
 in terms of the triplet and singlet scattering lengths is roughly
$2.3\times  10^{-23}~{\rm cm}^{2}$.
Once the cross-sections of the processes involved are known, the
diffusion scale is simply given by 
$L(\tau_{U})\sim \sqrt{6 \tau_{U} D(\tau_{U})}$, where $D(\tau_{U})$
is the diffusion coefficient (usually related to the mobility through
the Einstein coefficient \cite{63}) at any given time $\tau_{U}=M_{0}/T^2$.

At the onset  of nucleosynthesis ($T_{NS} \simeq 0.2 m_{e}$, where
$m_{e}$ is the electron mass) the comoving diffusion scale turns out
to be \cite{28,29,30,31} $3\times 10^{5} {\rm cm}$. The neutron diffusion
length blue-shifted at the time of the electroweak phase transition is
\begin{equation}
L_{diff}(T_{c}) = 0.3 ~{\rm cm}, ~{\rm for} ~~T_{c} =100~{\rm GeV}~.
\label{5.3}
\end{equation}
If $\Delta(L_{diff},t_{c}) > \frac{n_{B}}{s}$
matter--antimatter domains will not be erased by the nucleosynthesis
time and, at the same time,
fluctuations occurring over scales smaller than $L_{diff}(t_{c})$ at the
electroweak epoch are likely to be  dissipated \cite{28,29}. 

Taking again the flat spectrum for magnetic fields and assuming that
their energy is $\sim T^4$, we thus  obtain for the baryon-number fluctuations
at that scale $\delta(n_B/s) \sim 10^{-5} \gg 10^{-10}$.
If magnetic fields are large enough (with sufficiently flat spectra), 
domains of matter and antimatter may exist at scales 5 orders of
magnitude larger that the neutron diffusion
length \footnote{ Note that the energy fluctuations of the electroweak horizon
scale are always sufficiently small (i.e.  $\delta\rho_{p}/\rho_{p}
\ll 1$ for $r\sim L_{EW}$) so that  black-holes formation is not expected.}. 
To our best knowledge, there have been no studies of
non-homogeneous BBN with this type of initial conditions. Of course, there
were a lot of investigations of non-homogeneous nucleosynthesis,
motivated by first-order quark--hadron phase transition \cite{34}.  
In particular, baryon-number fluctuations with spectral  amplitudes
growing in frequency (and then decaying at large length scales) were
recently addressed \cite{30}, with the result that these
fluctuations are allowed, provided they occur at scales smaller that
the neutron diffusion length.
However,  Refs. \cite{28,29,30,31} essentially considered 
 positive-definite baryon number-fluctuations, rather than
with matter--antimatter domains. These results were also used  in order
to constrain the possible baryon-number fluctuations arising in the
context of topological defects models of baryogenesis \cite{54}.
It would be
very interesting to see whether matter--antimatter domains 
may change BBN bounds on the
baryon-to-photon ratio by changing the related predictions of the light
element abundances. This possible analysis will not be attempted here.

We can instead adopt  a more conservative attitude and require that no
matter--antimatter domains larger than the nucleon diffusion scale 
exist at the onset of nucleosynthesis. This
will give some constraints on the primordial hypercharge magnetic
fields. In order not to have matter--antimatter domains affecting BBN,
we therefore demand
\begin{equation}
\Delta(L_{diff},t_{c}) < \frac{n_{B}}{s}~~~.
\label{NSbound}
\end{equation}

\begin{figure}
\epsfxsize = 11 cm
\centerline{\epsffile{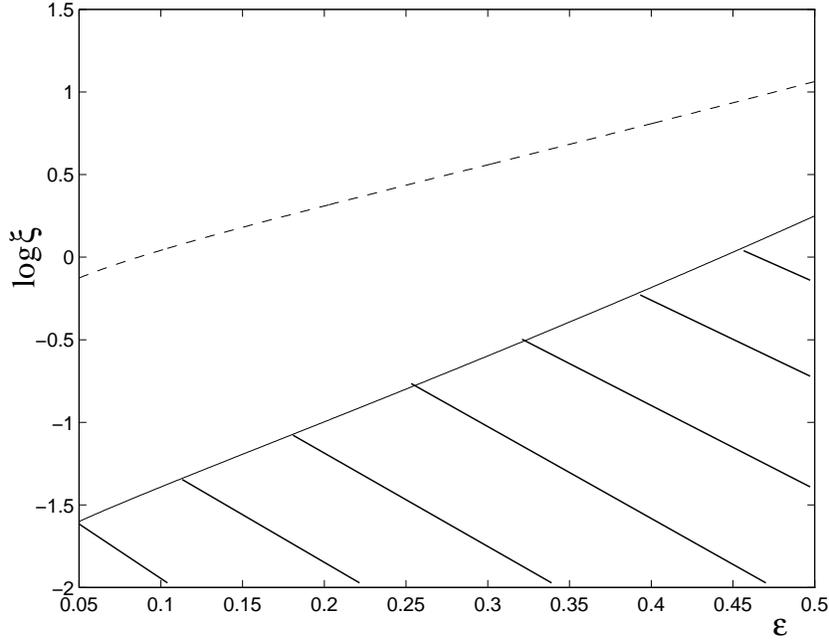}}
\caption[a]{We plot the constraint on the stochastic background of
magnetic hypercharge field derived in Eq. (\ref{exclNS})
 by requiring the homogeneity in the
baryon-number fluctuations at the neutron diffusion length. 
 We have chosen $\Omega_{B}h^2_{100} =0.01$ and
$\Omega_{0}h^2_{100}\sim 1$. We notice that by changing $h_{100}$ in the
range $0.4 <h_{100}<1$ the quantitative change in the bound is
negligible. We also took $\sigma_{c}/T_{c}\sim 70$ as fiducial value. The
variation of $\sigma_{c}$ in a plausible range does not alter the
features of the present plot.  The shaded region in 
the parameter space corresponds to matter--antimatter fluctuations that
will be erased by the nucleosynthesis time, whereas in order to have
sizeable matter--antimatter domains at the onset of BBN we have to go
above the line. In the dashed line we report the critical energy
density bound given in Eqs. (\ref{critbound}) and (\ref{critbound2}).}
\label{FIGU1}
\end{figure}

\begin{figure}
\begin{center}
\begin{tabular}{|c|c|}
      \hline
      \hbox{\epsfxsize = 7.5 cm  \epsffile{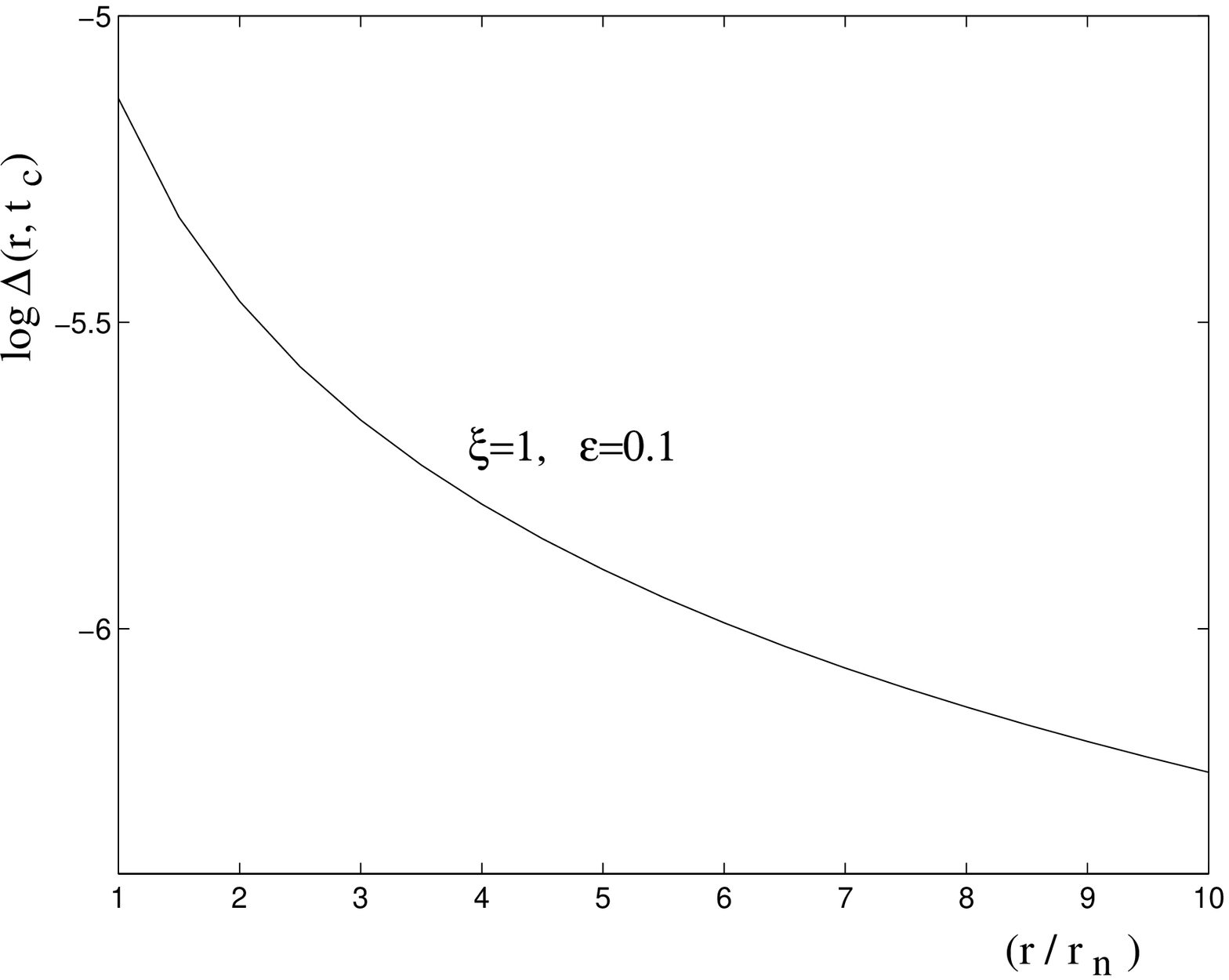}} &
      \hbox{\epsfxsize = 7.5 cm  \epsffile{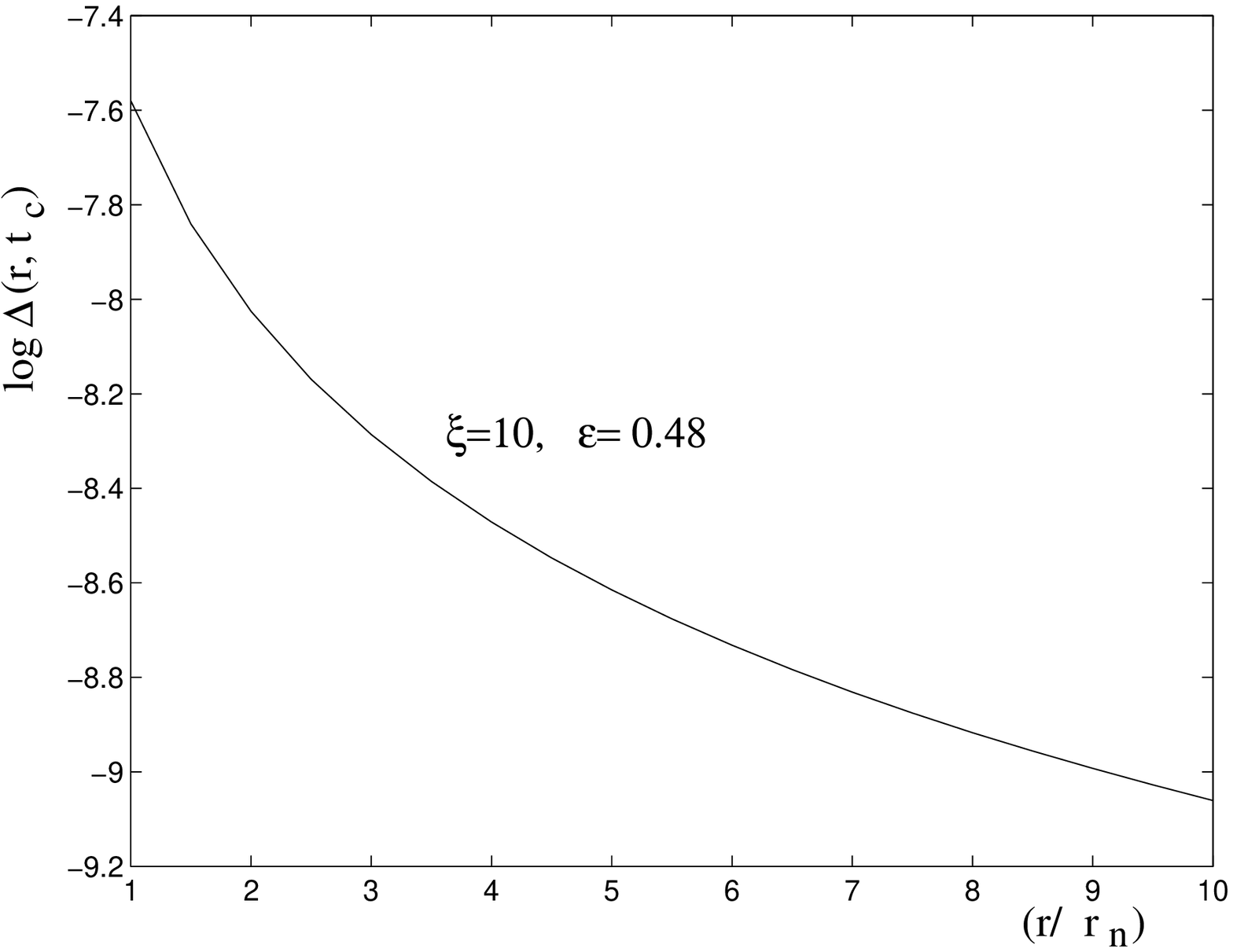}} \\
      \hline
      \hbox{\epsfxsize = 7.5 cm  \epsffile{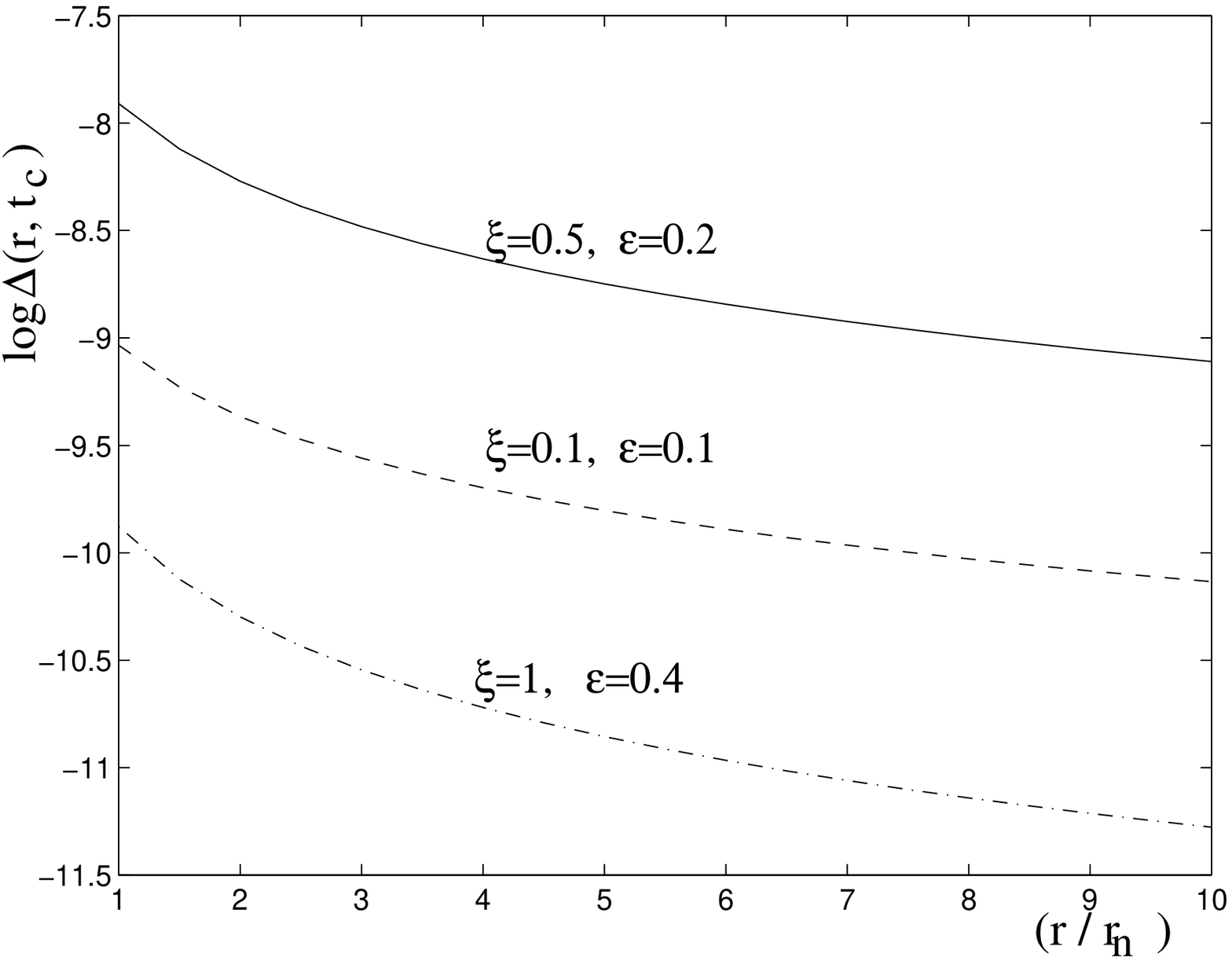}}  &
      \hbox{\epsfxsize = 7.5 cm  \epsffile{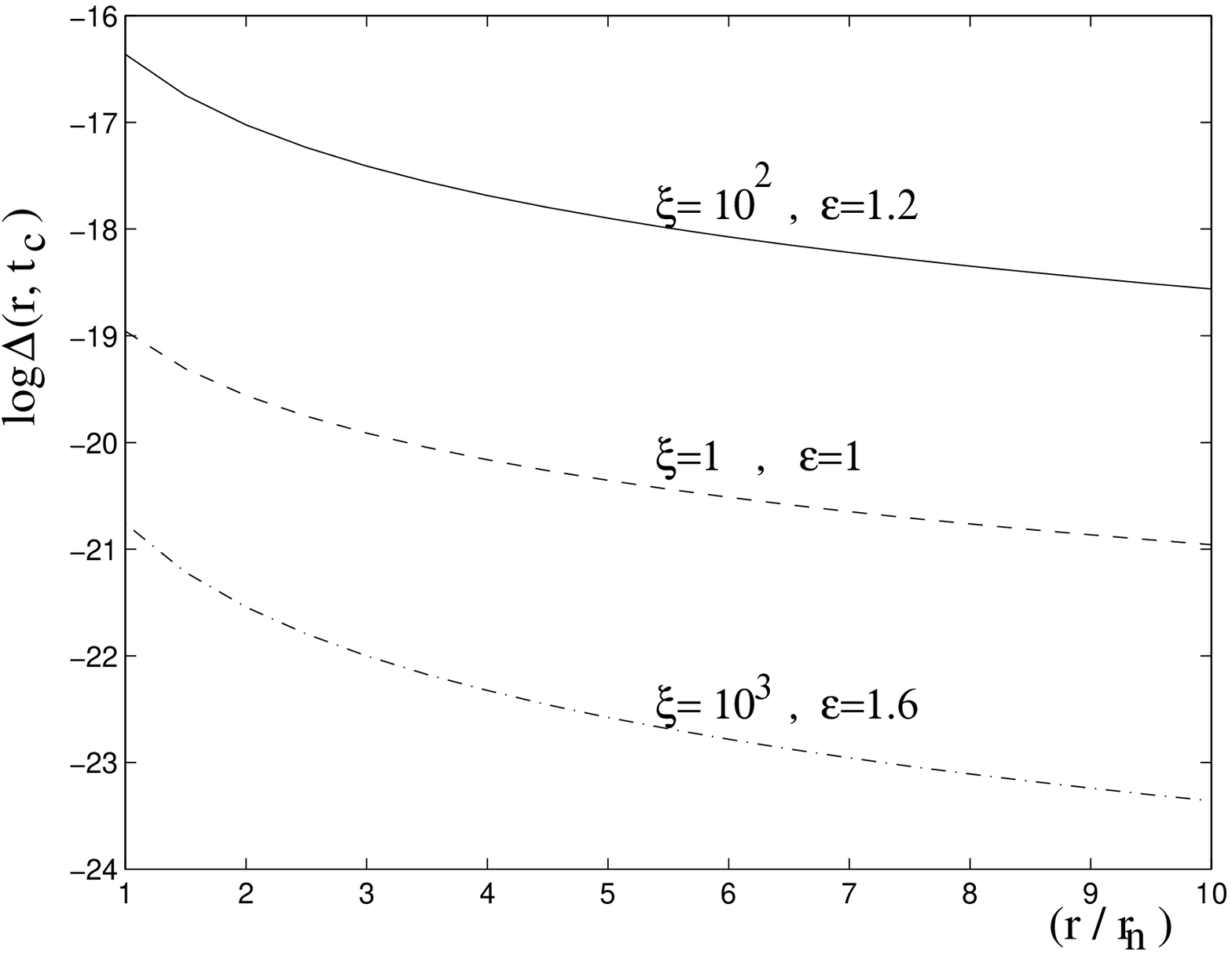}}\\
      \hline
\end{tabular}
\end{center}
\caption[a]{We plot the expected level of fluctuations given in
Eq. (\ref{Delta}) for scales of the order of (and larger than) the
neutron diffusion distance given in Eq. (\ref{5.3}). In the plot
$r_{n}= L_{diff}$. In this case we always took $\sigma_{c}/T_{c}\sim 70$. We
can notice that for flat enough spectra a very interesting level of
fluctuations is allowed. For violet spectra the fluctuations are
certainly suppressed. The same trend exists for even larger scales.}
\label{largescalefluct}
\end{figure}

From Eq. (\ref{Delta}), imposing the bound (\ref{NSbound}), we can translate the
constraints coming from homogeneity of BBN into  an exclusion plot in
 terms of the only two parameters ($\epsilon$ and $\xi$)
 characterizing our stochastic hypermagnetic background:
\begin{equation}
\log{\xi} < \left(-6.262 +\log{\frac{\sigma_c}{T_{c}}}  
+ \frac{1}{2}\log{\epsilon} 
+( 14.88)\epsilon +\log{[\Omega_{B} h^2_{100}]}\right)/(4 - \epsilon)~.
\label{exclNS}
\end{equation}
This condition is reported in {\bf Fig. \ref{FIGU1}} where, for the
validity of our approximations, we considered the case where
$\epsilon ~\gaq~0.05$ (for $\epsilon ~\laq ~0.05$ and scales
 $L_{diff}/r_{\sigma}
\gaq 10^{9}$ the corrections appearing in Eq. (\ref{Delta}) are no under
control). 
We plotted our bounds in the case $0.05~ \laq ~ \epsilon ~\laq ~1$. 
There is a second  constraint which one might want to impose on our
background, namely the one coming from the critical energy density:
\begin{equation}
\rho_{H}(t_{c}) <\rho_{\gamma}(t_{c}),~~~
\rho_{H}(t_{c})= \frac{1}{2} \langle|\vec{H}(\vec{x})|^2\rangle,
~~~\rho_{\gamma}(t_{c})=\frac{\pi^2}{30}N_{eff}T^4_{c}
\label{critbound}
\end{equation}
(where $ \langle|\vec{H}(\overline{x})|^2\rangle$ is given by 
Eq. (\ref{4.5}). Using now Eq. (\ref{energy}) we can convert 
(\ref{critbound}) into a further (but milder) constraint on our
parameter space
\begin{equation}
\log{\xi} < \left(\log{\frac{\pi~N_{eff}}{120}}
+\frac{\epsilon}{2}[\log{2} - \log{\frac{\sigma_{c}}{M_{0}}}] 
+ \log{\epsilon}\right)/(4-\epsilon) 
\label{critbound2}
\end{equation}
This bound is also reported in {\bf Fig. \ref{FIGU1}} (dashed
line).  
Sizeable matter--antimatter domains are produced when the 
spectrum is sufficiently flat.
This feature can also be deduced from the level of fluctuations for
scales of the order of the neutron diffusion distance, which we plot
 in {\bf Fig. \ref{largescalefluct}}. We see that it
is quite possible to get $\Delta \gg 10^{-10}$ around the neutron
diffusion distance.

We want  to notice that an artificial way of  relaxing
 our exclusion plot (\ref{exclNS}) could be 
(trivially) to
enhance the level of the baryon asymmetry by enhancing
 $\Omega_{B}h^2_{100}$ (up to values of the order of $0.1$-$1$). This 
phenomenon turns out to be  similar to the one discussed in
\cite{53}, where it was argued that baryon-number
fluctuations with blue frequency spectra might offer an interesting
mechanism for accounting for large amounts of baryonic dark matter.
This is purely an analogy, since the problem we are discussing (as we
stressed) is not the origin of the baryon asymmetry, but the possible
bounds on the hypercharge fields. Therefore, the value of  
$\Omega_{B}h^2_{100}$ is an external parameter for us, but not a 
computable number (see also Section 6).

A comment  is now in order concerning the phenomenological
estimates we made in this subsection.
The system of Eqs. (\ref{flatdiffusivity}) and (\ref{flatkinetic})
was solved in the approximation of local thermal equilibrium and the
possible back-reaction effects were ignored. Numerical solutions of
this non-linear system of partial differential equations are 
required if the level of induced fluctuations gets  too large
(i.e. $\Delta \sim 1$). We will not discuss here how to address this
complicated numerical problem, but we will come back to it in Section 6.

\subsection{Comparison with other bounds on primordial magnetic fields}

In this section we are going to compare the BBN bound 
reported in Eq. (\ref{exclNS}) with
 other possible bounds, which could be applied on primordial magnetic
 fields. 

It is well known that there are direct bounds on 
primordial  magnetic fields at  the nucleosynthesis epoch 
\cite{20}. 
Moreover, quite recently, two  constraints on magnetic field intensities
were derived using, respectively, the anisotropies in the microwave sky
 \cite{64} and the Faraday rotation
correlations \cite{65}. These last two bounds apply of course to the
case of fields that are, today, coherent over length scales much broader than
the (present) nucleosynthesis scale. 
In order to see how stringent our bound is, we should compare it with
the ones already available and coming from larger scales. In this
sense our aim is to show that our bounds are clearly more stringent
for small-scale fields but cannot compete (at even larger scales)
with the ones coming from the CMBR anisotropies and from the Faraday
rotation measurements.

Since the CMBR is isotropic to a very high degree of accuracy, its
small anisotropies can constrain the intensity of a constant
magnetic field (coherent over the present horizon size \cite{64}), which
could modify the evolution equations of the
matter sources by introducing a slightly anisotropic pressure  \cite{66}.
The calculation of the CMBR
anisotropies can be carried out also in the case of slightly
skew stresses, whose numerical weight depends upon the magnetic field
intensity. By comparing the final result with the level of
anisotropies detected by  COBE,  it is possible to compute
how big the magnetic field intensity should be not to
conflict with observed anisotropies in the microwave sky. At the
present time, the constraint is $|\vec{H}_{0}(t_{0})| 
< 6.8 ~\times~ 10^{-9} ( \Omega_{0} h^2_{100})^{1/2}~{\rm gauss}$ over
a length scale $L_{0}(t_{0}) \simeq 9.25\times 10^{27} ~
h^{-1}_{100} ~{\rm cm}$.
The authors of Refs. \cite{64,66} gave the bound in terms of $h_{50}$
(the present uncertainty in the Hubble parameter in units of $50~ {\rm
km}/{\rm Mpc ~sec}$). For consistency with our notation and in order to
make the comparison with other bounds easier, use instead 
$h_{100}=h_{50}/2$ taking, as usual, $0.4 < h_{100} < 1$. 
By blue-shifting the bound of \cite{64} up to $T_{c}$ we get
$|\vec{H}_{0}(t_{c})| 
< 1.12\times 10^{22} (\Omega_{0} h^2_{100})^{1/2}~ {\rm gauss}$ at a
scale $L_{0}(t_{c}) \simeq 7.19\times 10^{12}~h^{-1}_{100}~{\rm cm}$,
where  we assumed that the magnetic
field scales as $1/a^2(\tau)$, as it is plausible to demand  for length
scales larger than the magnetic diffusivity scale. 
We also took into account the evolution
in the effective number of degrees of freedom in the plasma, which is
$N_{eff}(t_{c}) \sim 106.75$ and $N_{eff}(t_{dec})\sim 3.90$.
The bound on $H_{0}(t_c)$ turns into a bound on $B(t_{c})=
\sqrt{\langle H^2\rangle}/T^2_{c}$ (recall that $1~{\rm gauss} = 1.95
\times ~10^{-20} {\rm GeV}^2$). 

\begin{figure}
\epsfxsize = 11 cm
\centerline{\epsffile{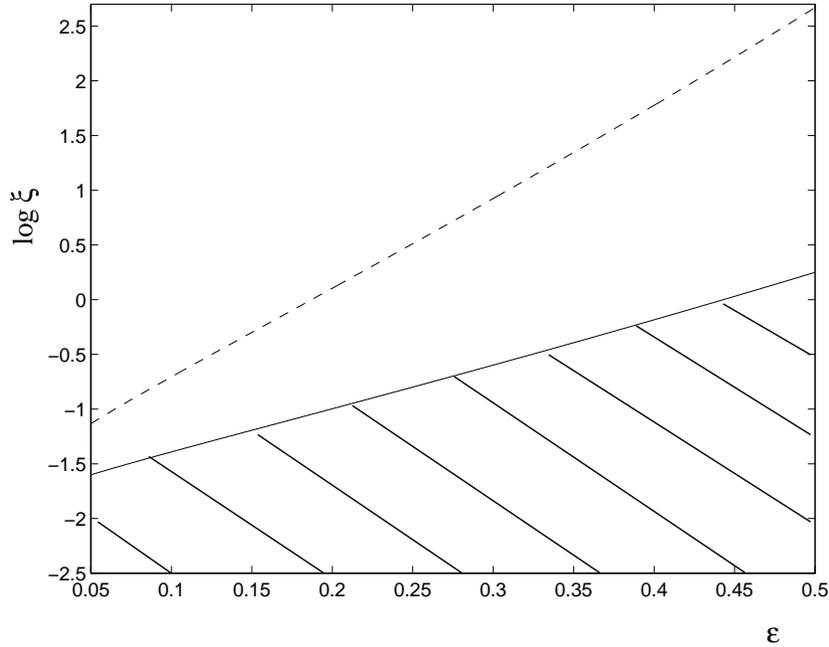}}
\caption[a]{We plot our bound (lower curve)  obtained in Eq. (\ref{exclNS})
and the bound derived in \cite{64,66} (upper curve)
 from the CMBR anisotropies in
the case of a magnetic field coherent over the horizon size today.
 For values of $\epsilon$ and $\xi$ above the shaded region 
  the level of fluctuations exceeds the bound 
(\ref{NSbound}). Again the variation of
$\Omega_B h_{100}^2$ (possibly between $0.1$ and $0.01$)
 changes the plot by only a few percent. In this plot we took
$h_{100}=0.6$.}
\label{FIGU3}
\end{figure}

\begin{figure}
\epsfxsize = 11 cm
\centerline{\epsffile{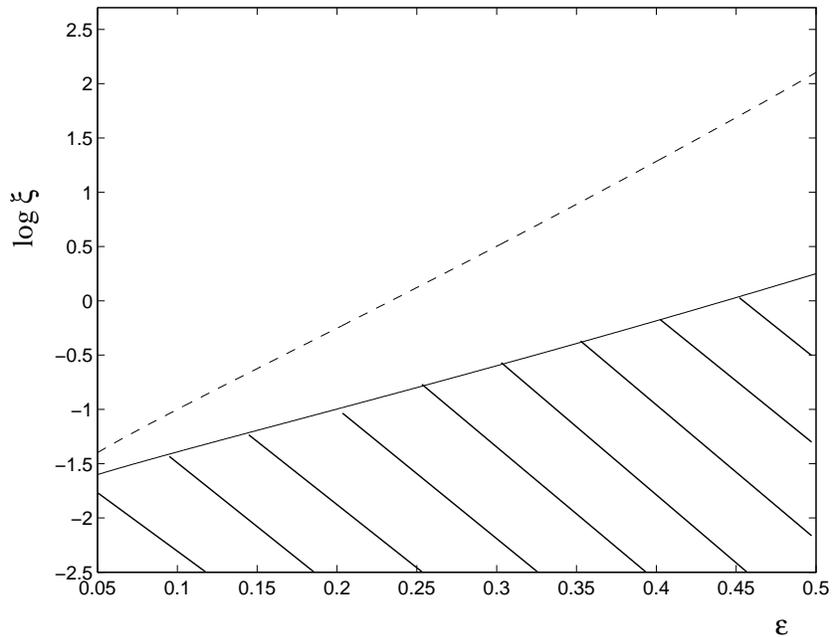}}
\caption[a]{We compare the bounds of Eqs. (\ref{exclNS}) and 
(\ref{5.19}). The most constraining bound is the
one given by the lower curve given by Eq. (\ref{exclNS}).
The upper curve derived in the context of Faraday rotation 
measurements is more
constraining  than the one obtained from the isotropy of the CMBR. 
The numerical value of the parameters for which this plot is
obtained is the same as for {\bf Fig. \ref{FIGU1}} and {\bf \ref{FIGU3}}.}
\label{FIGU4}
\end{figure}

\begin{figure}
\epsfxsize = 11 cm
\centerline{\epsffile{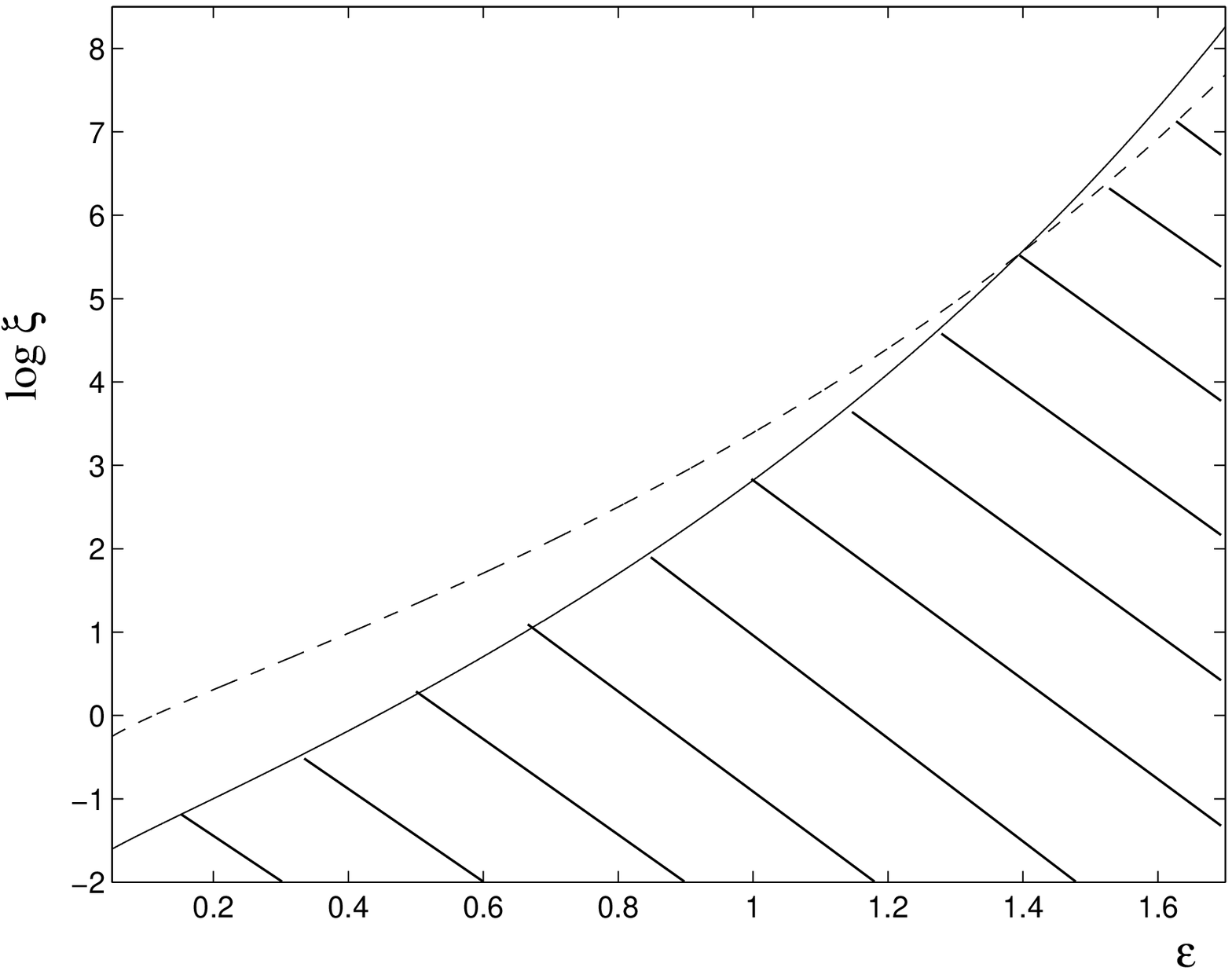}}
\caption[a]{We compare the direct bound on primordial magnetic fields at
the nucleosynthesis epoch derived in Eq. (\ref{NS})  with the bound
derived in Eq. (\ref{exclNS}), which applies to primordial magnetic
fields. We see that our
requirement is again more constraining than the one given in Refs. \cite{20}
for blue spectra,
whereas for $\epsilon~\gaq~1.4$ the bound given by (\ref{NS}) is better.}
\label{FIGU5}
\end{figure}

From Eq. (\ref{energy}) (taking $r\sim
L_{0}(t_{c})$, $T\sim T_{c}$)  it actually results, that the
parameter space of our model has to satisfy the following requirement
\begin{equation}
\log{\xi} < \left(-2.30 +\frac{1}{2}\log{[\Omega_{0}h^2_{100}]} 
 + \frac{1}{2}~ \log{\epsilon} +\bigg[ 14.13 -\frac{1}{2}\log{h_{100}}\bigg]~
\epsilon\right)/\left(2 - \frac{\epsilon}{2}\right).
\label{5.15}
\end{equation}
For comparison this constraint is reported in {\bf Fig.{\ref{FIGU3}}},
 together 
with the bound of {\bf Fig.{\ref{FIGU1}}}.
Since the region defined by Eq. (\ref{exclNS})  is always below the 
curve of Eq. (\ref{5.15}), we conclude  that the bound imposed
by the homogeneous BBN is more constraining 
than the one reported in \cite{64,66} for $\epsilon > 0.05$. Clearly,
for some very small slopes, the COBE bound will become better, but we
cannot compute this critical value of $\epsilon$ since our
approximation breaks down before it.
In \cite{65} the polarized emission
of few hundred galaxies was reviewed and another bound on the present
intensity of large-scale magnetic fields was reported. 
A large-scale field should
produce an additional shift in the polarization plane of the incoming
radiation according to the Faraday effect. 
In fact  the polarization plane of the
synchrotron radiation gets shifted, in the background of a magnetic
field, by an amount that is directly proportional to the integral of
the magnetic field times the electron density along the line of
sight. By subtracting, from the total angular 
 shift, the one  produced by each galactic field,
it is possible to constrain the intensity of any field coherent over
scales larger than the galactic one. The only uncertainty with this
procedure is that the measurements assume that the
magnetic fields of the Milky Way and of the other galaxies are known
to a very high degree of accuracy, since they have to be subtracted
from each estimate of the Faraday rotation. The  constraint
obtained with this method turns out to be 
$|\vec{H}_{1}(t_{0})| < 1\times [ \frac{2.6\times 10^{-7} {\rm
cm}^{-3}}{{\overline{n}}_{B}}] h_{100} \times 10^{-9} ~{\rm gauss}$,
for fields now coherent over scales 
$L_{1}(t_{0})\simeq (10-50) h^{-1}_{100}~{\rm Mpc}$.
 Assuming that the mean (present)
 baryon density is ${\overline{n}}_{B} \sim~1.13 \times
 10^{-5} (\Omega_{B} h^2_{100}) ~ {\rm cm}^{-3}$, 
the (blue-shifted) field intensity and its coherence length 
 will be respectively 
$|\vec{H}_{1}(t_{c})| < 3.79~ \times 10^{19}~(\Omega_{B} h^2_{100})^{-1} 
h_{100}~{\rm gauss}$ and
$L_{1}(t_{c}) \simeq  6\times 10^{10} h^{-1}_{100}~{\rm cm}$
where we took ( $L_{0}(t_{0}) \simeq 25~{\rm Mpc}$). 
By translating  this bound in the ($\xi,\epsilon$) plane, we
obtain the following relation
\begin{equation}
\log{\xi} < \left( - 4.77 + \frac{1}{2} \log{\epsilon}
-\frac{1}{2} \log{\Omega_{B}h^2_{100}} 
+ \bigg[13.09 -\frac{1}{2}\log{h_{100}}\bigg]~\epsilon\right)
/ \left(2 - \frac{\epsilon}{2}\right).
\label{5.19}
\end{equation}
This constraint is illustrated in {\bf Fig. \ref{FIGU4}}, where
 we see that our curve is
always below the curve reported in Eq. (\ref{5.19}). Our constraint is
again more stringent than the one of Eq. (\ref{5.19}) for $\epsilon > 0.05$.
The bounds on ordinary magnetic fields at the nucleosynthesis epoch
also apply in our case.
In order
not to affect  the universe expansion at nucleosynthesis it should hold
(see for instance Kernan et al. in Ref. \cite{20})
that $\rho_{H}<~0.27~\rho_{\nu}$ (where $\rho_{H}$ is
the magnetic energy density
defined in Eq. (\ref{critbound})) and $\rho_{\nu}$ is the energy density
contributed by the standard three light neutrinos for $T\ll 1~{\rm
MeV}$). Therefore in terms of $\xi$ and $\epsilon$ this bound reads:
\begin{eqnarray}
\log{\xi} < \frac{(11.30 -\frac{1}{2}\log{\frac{\sigma_{c}}{T_{c}}})\epsilon
+\log{\epsilon} - 0.2}{4 - \epsilon}~.
\label{NS}
\end{eqnarray}
This bound is  reported in {\bf Fig \ref{FIGU5}}
 and is compared with our bound.
We see that
the bound coming from the absence of matter--antimatter domains at the
nucleosynthesis epoch is more constraining for (by two orders of
magnitude for logarithmic energy spectra with  $\epsilon\ll 1$). Note
also that, according to \cite{Jed},  the bound (\ref{NS}) may  in fact
be absent, because there are other mechanisms, besides magnetic
diffusivity, that can dilute the magnetic fields before the
BBN. Then our bound remains the only one that can be applied to the
small-scale magnetic fields. 
\begin{figure}
\epsfxsize = 11 cm
\centerline{\epsffile{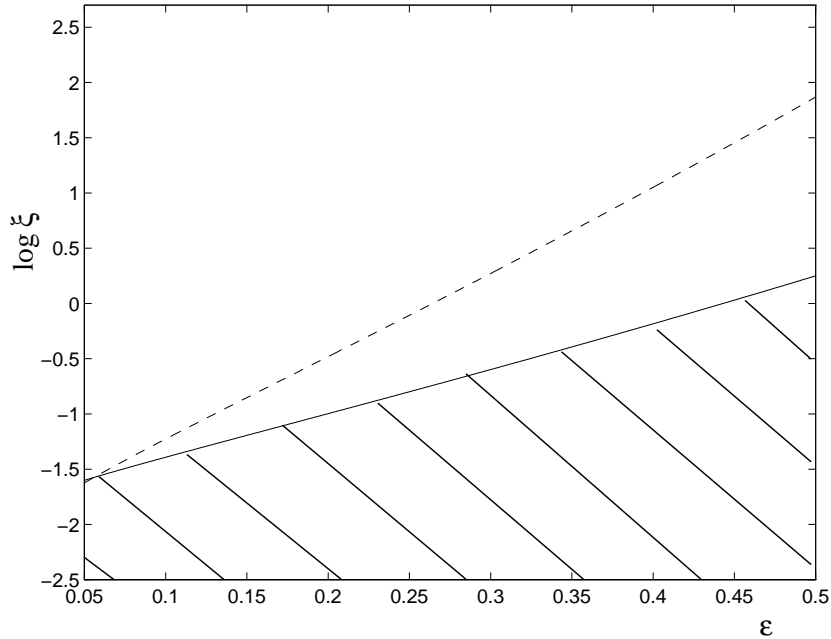}}
\caption[a]{We plot the constraint derived on the basis of BBN
considerations (Eq. (\ref{exclNS})) together with the requirement
derived in Eq. (\ref{faraday}). The dashed line corresponds to a
magnetic field of EW origin strong enough to rotate the polarization
plane of the CMBR under the assumption that the CMBR is polarized. We
can see that if such a field has an electroweak origin, then also sizeable
matter--antimatter fluctuations will be present for scales of the order
of the neutron diffusion distance. }
\label{faradayplot}
\end{figure} 

Another interesting numerical value of the magnetic field, which might
be compared with our considerations is$|\vec{H}_{3}(t_{dec})| 
\simeq 10^{-3}~ {\rm gauss}$ coherent over a scale 
$L_{3}(t_{dec}) \simeq 1.690\times 10^{23}~(\Omega_{0} h^2_{100})^{-1/2}~
{\rm cm}$, which is the size of the horizon at the decoupling.
If this field would be present at the decoupling time (when the
temperature was roughly $T_{dec}= 0.258~~{\rm ev}$) it might also rotate the
polarization plane of the CMBR provided the CMBR is weakly polarized
\cite{68}.
At the scale of the electroweak phase transition the blue-shifted
field and the corresponding correlation length read respectively
$|\vec{H}_{3}(t_{c})| \simeq 1.36\times 10^{21} ~{\rm gauss}$ and
$L_{3}(t_{c}) \simeq  1.447 \times 10^{11} (\Omega_0h^2_{100})^{-1/2}~
{\rm cm}$ and this imposes for our parameters the following requirement
\begin{equation}
\log{\xi} \gaq \left(-3.22 +\frac{1}{2} \log{\epsilon} + 13.28~\epsilon
-\frac{\epsilon}{4}\log{[\Omega_{0}h^2_{100}]}\right)/\left(2 -
\frac{\epsilon}{2}\right).
\label{faraday}
\end{equation}
It is of some interest to notice from  {\bf Fig. \ref{faradayplot}}
  that the region defined by
Eq. (\ref{faraday}) falls in the forbidden area of {\bf Fig. \ref{FIGU1}}. 
This means that the BBN bound of Eq. (\ref{exclNS}) excludes the
possibility that a primordial magnetic field of EW origin is strong
enough to rotate the polarization plane of the CMBR. 
On the other hand, if non-standard initial conditions for the
inhomogeneous BBN scenario (i.e. matter--antimatter domains)
 would be allowed, this conclusion might be
relaxed and the existence of such an intense field at the decoupling
epoch might be accommodated. It is anyway amusing that in our present
discussion the existence of a magnetic field at the decoupling epoch
might imply the presence of small-scale antimatter domains at the
onset of BBN.

\subsection{The rate of right-electron chirality flip}

In Section 3 we pointed out how important is, in our context, the
interplay between the ``perturbative'' rate given by the right electron
chirality flip processes and the ``non-perturbative'' one coming from
the anomaly. For the MSM the perturbative rate of
chirality flip has been computed in \cite{42} and is determined by the
right electron Yukawa coupling. If MSM is a correct theory, then
$\Gamma > \Gamma_{\cal H}$ only for extremely small magnetic fields,
$H^2/T^4 < \frac{22}{783} \frac{\sigma_0}{\alpha'}\frac{T_R}{M_0}
\simeq 10^{-11}$. So weak hypermagnetic fields do not produce any
interesting fluctuations. For larger magnetic fields the
approximation outlined in Eq. (\ref{2.26}) must be used.
 The amplification factor that appears in
Eq. (\ref{final}) can be extracted from  \cite{42}:
\begin{equation}
\int_{0}^{t_{c}} d t \Gamma \simeq \frac{\Gamma M_{0}}{T_{c}^2} \simeq 350
 \left( \frac{100 {\rm~GeV}}{T_{c}}\right) [ (-1.1+ 2.4~x_{H}) + 1.0 
+ h_{t}^2 ( 0.6 - 0.09~ x_{H})],
\label{2.33}
\end{equation}
where $x_{H} = \lim_{T\rightarrow \infty} \frac{m_{H}(T)}{T}$
is the high-temperature limit of the Higgs thermal mass and $h_{t}$ is
the top-quark Yukawa coupling. Taking $x_{H}\simeq 0.6$ (for which
the scattering contribution to the rate is always dominant with resepct to
the decay contribution), we find that the integrated rate is 
$655 \times~ (100~{\rm GeV}/T_{c})$. With the use of this number the
analysis of Section 5 can be redone with the result that no
interesting baryon-number fluctuations can be produced from stochastic
hypermagnetic background. So, for MSM, one hardly expects any 
cosmological consequences coming from
the background of the type (\ref{4.4}) (for other types of primordial
hypermagnetic fields, 
considered below in Section 6.2, the conclusion is different).   

However, in the extensions of the standard model, the rate $\Gamma$ 
can be naturally larger than in the MSM. For example, in the context of the 
Minimal Supersymmetric Standard Model (MSSM) 
the right-electrons Yukawa coupling is enhanced
 by a factor $1/\cos{\beta}$ ($\tan\beta$ gives the ratio of the
expectation values of the two doublets), so that $T_{R}$ can
be larger by a factor $10^{3}$ for experimentally allowed values
of $\tan\beta\sim 50$.
Moreover  the right-electron number
is now shared between electrons and selectrons,  and it is necessary 
to consider also processes that change the selectron number.

The question we now  want to address is more phenomenological. 
Namely we want to see how large the perturbative chirality flip rate 
$\Gamma$ should be in order to produce 
sizeable matter--antimatter fluctuations, which could influence the
BBN. For this purpose,  we just require that $\Gamma > \Gamma_{\cal H}$,
with $\Gamma_{\cal H}$ taken from  Eqs. (\ref{final}) and (\ref{energy}),
and use the minimal amplitude of the hypermagnetic field obtainable
 from the bound
(\ref{exclNS}), which can produce sizeable matter--antimatter
fluctuations. This gives $\Gamma/T_c > 10^{-9}$, which corresponds to
the right electron perturbative freezing temperature $T_R = 10^5$ TeV.
As we discussed above, these values of the temperature is perfectly
possible, say, in the MSSM.
If the actual value of the freezing
temperature is smaller than $10^5$ TeV, the stochastic hypercharge
background of type (\ref{4.4}) produces  baryon-number
fluctuations  too small to affect BBN.
It is then interesting that a quite energetic stochastic
hypermagnetic background can be accommodated in the MSM without any
(potentially dangerous) implications. The energy density stored in
this background can then be able to influence the dynamics of the EW
phase transition without conflicting with any bound derived from
BBN. This will be one of the subjects of the following section.

\renewcommand{\theequation}{6.\arabic{equation}}
\setcounter{equation}{0}
\section{EW phase transition and baryogenesis}

The aim of this section is the discussion of the influence of the
hypercharge magnetic field on the electroweak baryogenesis (for 
reviews, see \cite{rev}). First, we will consider the EW phase
transition in the presence of the hypermagnetic field. Then, we will
show that the occurrence of some specific
hypermagnetic configurations in the EW plasma could be responsible for 
the baryon asymmetry of the Universe (BAU).

\subsection{EW phase transition}
The hypercharge magnetic field, present at high temperature,  
can influence the dynamics of the phase transition. The physical
picture  of this phenomenon is exactly the same as 
the macroscopic description of  superconductors in the
presence of an external magnetic field. The normal-superconducting phase
transition, being of second order in the absence of magnetic
fields, becomes  of first order if a  magnetic field is present. The reason
for this is the Meissner effect, i.e.  the fact that the magnetic field cannot
propagate inside a superconducting cavity, and, therefore, creates
 an extra pressure acting on the normal-superconducting boundary \cite{LL}. 
Our consideration below explores this simple picture.

Consider the plain domain wall that separates broken and symmetric
phase at some temperature $T$, in the presence of a uniform 
hypercharge magnetic field ${\cal Y}_{j}$. Far from the domain wall, in
the symmetric phase, the non-Abelian SU(2)  field strength 
(${\cal W}_{j}^{3}$) is equal to
zero, because of a non-perturbative mass gap generation. Inside the
broken phase, the massive $Z_{j}$ combination of ${\cal Y}_{j}$  and ${\cal
W}^{3}_{j}$,
\begin{equation}
Z_{j} = \cos{\theta_{W}} {\cal W}_{j}^{3} - \sin{\theta_{W}}{\cal Y}_{j}
\end{equation} 
must be equal to zero, while the massless combination, corresponding
to photon $A_{j}^{em}$, survives. The matching of the fields on the
boundary gives $A^{em}_{j} = {\cal Y}_{j} \cos{\theta_{W}}$. Thus, an
extra pressure $\frac{1}{2}|\vec{{\cal H}}_{Y}|^2\sin^2\theta_{W}$
acts on the domain wall from the symmetric side. At the critical
temperature it must be compensated by the vacuum pressure of the
scalar field. If we neglect loop corrections associated with the
presence of magnetic fields, then the condition that determines the
critical temperature is:
\begin{equation}
\frac{1}{2}|\vec{{\cal H}}_{Y}|^2\sin^2\theta_{W} = 
V(0,T_c) - V(\varphi_{min},T_c)~~,
\label{pressure}
\end{equation}
where $ V(\varphi,T)$ is the effective potential in the absence of
magnetic field, $\varphi_{min}$ is the location of the minimum of the
potential at temperature $T$.

The above consideration was dealing with the uniform magnetic
fields. Clearly, it remains valid when the typical distance scale of
inhomogeneities of magnetic field are larger than the typical bubble
size. This is the case for bubbles smaller than the magnetic
diffusion scale, and, in particular, at the onset of the bubble
nucleation. Thus, the estimate of the critical temperature coming from
(\ref{pressure}) is applicable. For bubbles larger than the
diffusivity scale, the presence of a stochastic magnetic field will
considerably modify their evolution. In particular, the spherical form
of the bubbles is very likely to be spoiled.

\begin{figure}
\epsfxsize = 11 cm
\centerline{\epsffile{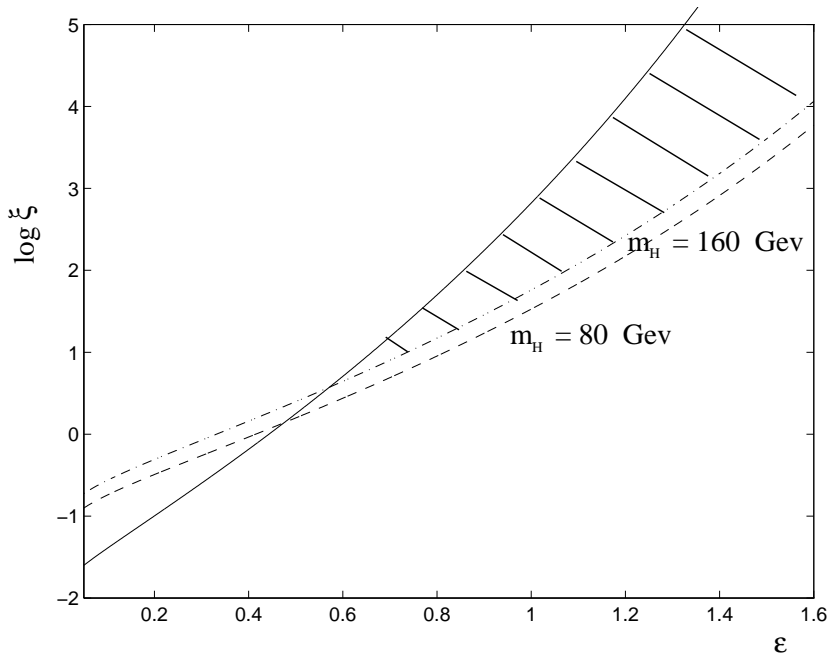}}
\caption[a]{We plot the requirement obtained in Eq. (\ref{pressure}) in
the case of a stochastic hypermagnetic background. The dashed line
refers to the case of $m_{H}= 80~{\rm GeV}$, whereas the dot-dashed
line refers to the case of $m_{H} = 160~{\rm GeV}$. With the full line
is reported for comparison the bound coming from BBN and discussed in
Eq. (\ref{exclNS}). We can clearly see (shaded region) that for steep enough
hypermagnetic energy spectra (i.e. $\epsilon~\gaq~0.4~$-$0.6$) it is
possible to have a strongly first-order EW phase transition
 consistent with the bound (\ref{exclNS}).}
\label{FIGU10}
\end{figure}

Relation (\ref{pressure}) may be used to define a minimum magnetic
field, which can make a phase transition  strong enough to allow
electroweak baryogenesis in the MSM. One can just fix the Higgs mass, find the
temperature at which the minimum of the effective potential satisfies
the constraint $\varphi_{min}/T >1$ \cite{25}, and read off the
hypermagnetic field from Eq. (\ref{pressure}). With the use of the
two-loop effective potential computed in \cite{PT}, we get, 
for  $m_{H} \sim 80~{\rm GeV}$, $\langle |\vec{{\cal H}}_{Y}|^2 
\rangle /T_{c}^4~\gaq ~0.06$, whereas if $m_{H} \sim 160~{\rm GeV}$ 
we will have 
 $\langle |\vec{{\cal H}}_{Y}|^2 \rangle /T_{c}^4 ~\gaq ~ 0.3$. 
For stochastic backgrounds these constraints are plotted in
{\bf Fig. {\ref{FIGU10}}} in terms of our variables $\xi$ and
$\epsilon$, characterizing the spectrum. We see that the region
of parameter space where a strongly first-order EW phase transition is
possible, without spoiling BBN with excessive matter--antimatter domains
extends, from $\epsilon\simeq 0.5$ up to
$\epsilon > 1$. Therefore we come to the conclusion that for violet
hypermagnetic energy spectra the level of induced fluctuations is
quite tiny at the neutron diffusion distance, but the dynamics of the
phase transition can be strongly affected. The magnetic fields, which
can modify the nature of the phase transitions do  not appear to be 
subjected to any other constraints.

The observation that the presence of primordial hypercharge magnetic
fields at the electroweak epoch may make an EW phase transition 
strongly first-order removes a main objection against the possibility of
baryon asymmetry generation in the MSM
\cite{25,rev}. We also
note, that the background magnetic field breaks C and CP
symmetries, which may considerably change the analysis of different
processes near the domain wall, which are used in EW baryogenesis
mechanisms. This study is beyond the scope of this paper. Instead, we
will point out in the next subsection that some specific
configurations of the hypercharge magnetic field may create the  net
baryon asymmetry. The following discussion  is similar to considerations of
the baryogenesis from Chern-Simons condensate in \cite{25}.

\subsection{Chern-Simons condensates and  the BAU}
In  Sections 4 and 5   we were concerned with the case of
stochastically distributed hypermagnetic fields. In this case, the
average baryon asymmetry vanishes 
(i.e. $\langle\delta(n_{B}/s)(\vec{x},t_{c})\rangle=0$) in spite of
the fact that the fluctuations of the baryon number may be considerable.
Thus, we assumed that the source of the baryon asymmetry has no relation to
the primordial hypermagnetic fields. For example, the BAU may have
been generated  because of GUT interactions or at the EW phase transition.

In this subsection we are going to discuss hypermagnetic backgrounds
that may give rise to the BAU. We have
no idea whether these types of background can or cannot be generated by
some mechanism. Our aim is to point out the essential features
of the hypermagnetic field that are necessary for the production of net 
baryon asymmetry.

If the hypermagnetic field configuration is topologically non-trivial
(i.e. $\vec{\cal H}_{Y}\cdot\vec{\nabla}\times\vec{\cal
H}_{Y}\neq 0$), then  $\delta(n_{B}/s)\neq 0$. 
As a particular example we will discuss the case of 
a Chern-Simons wave configuration
\begin{eqnarray}
{\cal Y}_{x} &=& {\cal Y}(t) \sin{k_{0} z},
\nonumber\\
{\cal Y}_{y} &=& {\cal Y}(t)  \cos{k_{0} z},
\nonumber\\
{\cal Y}_{z} &=& 0~.
\label{2.39}
\end{eqnarray}
The hypermagnetic field is 
$\vec{\cal H}=\vec{\nabla}\times{\cal Y}$
and the magnetic helicity is simply
\begin{eqnarray}
\overline{{\cal H}}\cdot \overline{\nabla} \times {\overline{\cal H}},
= k_{0} {\cal H}^2(t)
\nonumber
\end{eqnarray}
with ${\cal H}(t) = k_{0} {\cal Y}(t)$.
Thus from (\ref{2.39}) we obtain that:
\begin{eqnarray}
{\cal H}^2_{x} + {\cal H}^2_{y} + {\cal H}^2_{z}= {\cal H}^2 (t),~~~
n_{CS} = - \frac{g'^2}{32 \pi^2} {\overline{\cal
H}}\cdot{\overline{\cal Y}} = \frac{g'^2}{ 32 \pi^2 ~ k_{0}} {\cal
H}^2(t)
\nonumber
\end{eqnarray}
We notice that this configuration describes a magnetic knot with
uniform energy and Chern-Simons density over the whole observable
Universe. Similar configurations are used in the
MHD treatment of the dynamo instability \cite{2}.
Inserting the configuration (\ref{2.39}) into the 
 evolution equations (\ref{flatdiffusivity}) and into
(\ref{flatkinetic}), we get a system of non-linear
 ordinary differential equations
\begin{eqnarray}
\frac{d {\cal B}}{d w} &=&-\frac{4~\alpha'}{\pi} \frac{T}{\sigma_{c}}
\left(\frac{k_0}{T}\right)\left(\frac{\mu_{R}}{T}\right) {\cal B} -
\frac{T}{\sigma_{c}}\left(\frac{k_{0}}{T}\right)^2 {\cal B},~~~{\cal
B}(t)= \frac{{\cal H}(t)}{T^2},
\nonumber\\
\frac{d }{d w} \left(\frac{\mu_{R}}{T}\right) &=& 
-\left(\frac{\alpha'}{ \pi}\right)^2
\frac{783}{88}\left(\frac{k_{0}}{\sigma_{c}}\right){\cal B}^2 - 
\left(\frac{\Gamma}{T} +\frac{\Gamma_{{\cal B}}}{T}\right) 
\left(\frac{\mu_{R}}{T}\right)  , ~~~
{\Gamma_{{\cal B}}}
=\frac{\alpha'}{\pi}\frac{783}{22}\frac{T}{\sigma_{c}}{\cal B}^2 T
\label{2.40}
\end{eqnarray}
(where  $w = t T$).
Equations (\ref{2.40}) are in fact known as a  
generalized Lotka-Volterra system \cite{50}. They  can be solved
numerically for different types of initial conditions. As in the
discussion of the stochastic backgrounds,
 we will consider our system in the  adiabatic
approximation and we  will then  use  the  general set of  equations 
(\ref{flatdiffusivity}) and (\ref{flatkinetic}), valid
for arbitrary backgrounds.

As usual, the magnetic diffusivity $\sigma_{c}$ defines
the diffusion scale and therefore
a Chern-Simons wave configuration with  typical momentum ($k_{0}$) 
larger than $k_{\sigma}\sim 10^{-7} T_{c}$ will be washed out. For
smaller $k_0$, we get from (\ref{final}):
\begin{eqnarray}
\left(\frac{n_{B}}{s}\right)(\vec{x}, t_{c}) &\simeq & 
\frac{\alpha'}{2~\pi~\sigma_{c}} \left(\frac{n_{f}}{s}\right) 
\left(\frac{k_{0}}{T_{c}}\right) \left(\frac{M_{0}}{T_{c}}\right)
{\cal H}^2(t_{c})\simeq 
10^{10} \left(\frac{k_{0}}{T_{c}}\right) 
\left(\frac{ {\cal H}^2}{T^4_{c}}\right),~~{\rm for}~~~\Gamma \gaq \Gamma_{{\cal H}}
\nonumber\\
\left(\frac{n_{B}}{s}\right)(\vec{x}, t_{c}) &\simeq &
\frac{11 \pi }{783~\alpha'} \left(\frac{n_{f}}{s}\right)k_{0}
\Gamma M_{0}\simeq  0.3 
\left(\frac{T_{R}}{T_{c}}\right)
\left(\frac{k_{0}}{T_{c}}\right),~~~{\rm for}~~~~ \Gamma \laq 
\Gamma_{{\cal H}}.
\label{CSfluct}
\end{eqnarray}
Let us now assume that we work only in the framework of the MSM. Then, 
in order to produce baryon asymmetry we need a strong enough
first-order phase transition, and therefore, a strong enough magnetic field
(see previous subsection). Thus, $\Gamma_{{\cal H}}~\gaq~\Gamma$. Using
the fact that, in the MSM,  $T_{R}\simeq 80~{\rm
TeV}$ we see that $\left(\frac{n_{B}}{s}\right)\simeq 10^{-10}$ for
$k_0/T_c \simeq 10^{-12}$. This value is well below the magnetic
diffusivity scale and, therefore, this type of configuration, if ever
created, will survive till the EW epoch.

In the extensions of the standard model one may have a strong enough
electroweak phase transition without any magnetic field \cite{MSSM}. In
addition, the perturbative electron chirality rate may be considerably
higher than in the MSM. Thus, the hierarchy $\Gamma_{{\cal H}}~\laq
~\Gamma$ may be naturally realized. Then, for 
$ \left(\frac{k_{0}}{T_{c}}\right) 
\left(\frac{ {\cal H}^2}{T^4_{c}}\right)\simeq 10^{-20}$ the BAU
calculated from the hypermagnetic field is of the right
order of magnitude. For example if the typical momentum $k_{0}$
 of the Chern-Simons
condensate is of the  order  of the EW horizon at $T\sim T_{c}$
(i.e. $k_0 \sim L_{ew}^{-1}\simeq 10^{-16} T_{c}$) then for small
enough hypermagnetic energy (i.e. $ {\cal H}^2\simeq ~10^{-4}T_{c}^4$) the
BAU is $\sim 10^{-10}$.
Thus, it is not excluded  that the baryon asymmetry of the Universe is a
consequence of the topologically non-trivial primordial 
hypercharge magnetic field.
 
\renewcommand{\theequation}{7.\arabic{equation}}
\setcounter{equation}{0}
\section{Concluding remarks}

There are no compelling theoretical reasons against the
existence of long lived Abelian hypercharge fields at the electroweak
epoch. In the present paper we showed that, if they did exist,
they could have a number of cosmological consequences. The stochastic
hypermagnetic backgrounds induce baryon-number fluctuations because
of the electroweak anomaly. These fluctuations may produce sizable
matter--antimatter domains at the onset of BBN
and affect its  predictions. Magnetic
fields can change considerably the dynamics of the electroweak phase
transition in the MSM and make it strongly first-order even for 
large Higgs masses. Topologically non-trivial hypermagnetic
configurations may be responsible for the BAU.

We left aside a number of questions that may be subjects of
future studies. For example, in the study of AMHD equations, we
focused our attention on the case where the correlation scale of the
velocity field was much smaller than that of the
magnetic field, and we also assumed that the velocity field was
(globally) invariant under parity transformations (i.e. in the absence of
global vorticity over the EW horizon, at the epoch of the phase
transition). Owing to different phenomena (e.g. bubble collision)
turbulence  may arise inside the EW horizon, leading to a
 non-mirror symmetric velocity field over some length scale
typical of the mechanism responsible for the turbulence.
 If the turbulence produces a non-zero vorticity  of the
plasma, then the collective plasma motions may be transformed into
fermion number via the amplified hypermagnetic field through a kind
of EW dynamo mechanism. We completely neglected the possible
occurrence of (global) vorticity, and to relax this hypothesis may be 
interesting.

We do not know what is the possible influence of
matter--antimatter domains on the inhomogeneous BBN scenario. In
particular we have no idea if some spectral distribution  of  hypermagnetic 
fields could induce a spectrum of baryon-number fluctuations, which 
can lead to  a  larger baryon-to-photon ratio.

\section*{Acknowledgements}

We wish to thank J. Cline, M. Joyce, H. Kurki-Suonio and
G. Veneziano for interesting comments and helpful
discussions. 

\begin{appendix}

\renewcommand{\theequation}{A.\arabic{equation}}
\setcounter{equation}{0}
\section{Magnetic helicity correlations ($\Gamma> \Gamma_{H}$)}

The aim of this Appendix is to compute explicitly the two-point
correlation function of the magnetic helicity $\Lambda(\vec{x})$, namely
\begin{equation}
\langle \Lambda(\vec{x}) \Lambda(\vec{x}+\vec{r})\rangle =
\langle(\vec{H}\cdot\vec{\nabla}\times\vec{H})(\vec{x})(
\vec{H}\cdot\vec{\nabla}\times\vec{H})
(\vec{x}+\vec{r})\rangle
\label{B.1}
\end{equation}
in terms of the two-point function
\begin{equation}
G_{ij}(r) = F_{1}(r)\delta_{ij} + r_{i}r_{j} F_{2}(r).
\label{B.2}
\end{equation}
The results illustrated here are quite relevant for a
precise estimate of the level of fluctuations induced by a stochastic 
background of hypermagnetic fields. 

The estimate of the correlation function (\ref{B.1})
 may be carried out either in real space or in Fourier
space. In Fourier space the calculation can be reduced to the
estimate of a convolution, whereas in real space the main algebraic
task is to compute the various derivatives appearing in the ensemble
average.

The stochastic average appearing in Eq. (\ref{B.1})
 can be rewritten as 
\begin{eqnarray}
& &\langle(\vec{H}\cdot\vec{\nabla}\times\vec{H})
(\vec{x})(\vec{H}\cdot\vec{\nabla}\times\vec{H})
(\vec{y})\rangle=
\nonumber\\
& &\epsilon_{ijk}\epsilon_{mnl}
\lim_{\overline{x}'\rightarrow\overline{x}}
\lim_{\overline{y}'\rightarrow\overline{y}}\frac{\partial}{\partial x^{i}}
\frac{\partial}{\partial y^{m}} \langle H_{k}(\overline{x}')
H_{j}(\overline{x}) H_{l}(\overline{y}') H_{n}(\overline{y}) \rangle.
\label{B.3}
\end{eqnarray}
This expression turns out to be quite useful, since it allows us to
perform the derivations with respect to $x^i$ and $y^m$ after the
average is taken. 

If we now use the fact that the background of hypercharge fields is
stochastic, we can write that
\begin{eqnarray}
& &\langle H_{k}(\vec{x}')
H_{j}(\vec{x}) H_{l}(\vec{y}') H_{n}(\vec{y}) \rangle=
 [ \langle H_{k}(\vec{x}')H_{j}(\vec{x}) \rangle
\langle H_{l}(\vec{y}') H_{n}(\vec{y}) \rangle +
\nonumber\\
& & \langle H_{k}(\vec{x}') H_{l}(\vec{y}')\rangle
\langle H_{j}(\vec{x}) H_{n}(\vec{y})\rangle +
\langle H_{k}(\vec{x}') H_{n}(\vec{y})\rangle\langle
H_{j}(\vec{x}) H_{l}(\vec{y}') \rangle].
\label{B.4}
\end{eqnarray}
Inserting in Eq. (\ref{B.4}) the representation (\ref{B.2}), we get the
following relation:
\begin{eqnarray}
& &\langle(\vec{H}\cdot\vec{\nabla}\times\vec{H})(\vec{x})(
\vec{H}\cdot\vec{\nabla}\times\vec{H})
(\vec{y})\rangle=
\nonumber\\
& &\epsilon_{ijk}\epsilon_{mnl}
\lim_{\vec{x}'\rightarrow\vec{x}}
\lim_{\vec{y}'\rightarrow\vec{y}}\frac{\partial}{\partial x^{i}}
\frac{\partial}{\partial y^{m}}[ 
A(\vec{x},\vec{x}';\vec{y},\vec{y}')+
B( \vec{x},\vec{x}';\vec{y},\vec{y}')+
C(\vec{x},\vec{x}';\vec{y},\vec{y}' )],
\label{B.5}
\end{eqnarray}
where
\begin{eqnarray}
A(\vec{x},\vec{x}';\vec{y},\vec{y}')&=&[
F_{2}(|\vec{x}'- \vec{x}|)F_{2}(|\vec{y}'-
\vec{y}|)(x'_{k}-x_{k})(x'_{j}-x_{j})(y'_{l}-y_{l})(y'_{n}-y_{n})
\nonumber\\
&+&F_{1}(|\vec{x}'- \vec{x}|)F_{2}(|\vec{y}'-
\vec{y}|)\delta_{kj}(y'_{l}-y_{l}) (y'_{n}-y_{n}) 
\nonumber\\
&+&
F_{2}(|\vec{x}'-
\vec{x}|)F_{1}(|\vec{y}'-\vec{y}|)
\delta_{nl}(x_{k}' - x_{k})(x'_{j}-x_{j})
\nonumber\\
&+& F_{1}(|\vec{x}'- \vec{x}|)F_{1}(|\vec{y}'-
\vec{y}|)\delta_{kj}\delta_{ln}].
\nonumber\\
B( \vec{x},\vec{x}';\vec{y},\vec{y}')&=&[
F_{2}(|\vec{x}'- \vec{y}'|)F_{2}(|\vec{x}-
\vec{y}|)(x'_{k}-y'_{k})(x'_{l}-y'_{l})(x_{j}-y_{j})(x_{n}-y_{n})
\nonumber\\
&+&F_{1}(|\vec{x}'- \vec{y}'|)F_{2}(|\vec{x}-
\vec{y}|)\delta_{kl}(x_{j}-y_{j}) (x_{n}-y_{n}) 
\nonumber\\
&+&
F_{2}(|\vec{x}'-
\vec{y}'|)F_{1}(|\vec{x}-\vec{y}|)
\delta_{nj}(x_{k}' - y'_{k})(x'_{l}-y'_{l})
\nonumber\\
&+& F_{1}(|\vec{x}'- \vec{y}'|)F_{1}(|\vec{x}-
\vec{y}|)\delta_{kl}\delta_{jn} ],
\nonumber\\
C( \vec{x},\vec{x}';\vec{y},\vec{y}')&=&
[F_{2}(|\vec{x}'- \vec{y}|)F_{2}(|\vec{x}-
\vec{y}'|)(x'_{n}-y'_{n})(x'_{k}-y_{k})(x_{j}-y_{j}')(x_{l}-y'_{l})
\nonumber\\
&+&F_{1}(|\vec{x}'- \vec{y}|)F_{2}(|\vec{x}-
\vec{y}'|)\delta_{kn}(x_{j}-y_{j}') (x_{l}-y'_{l}) 
\nonumber\\
&+&
F_{2}(|\vec{x}-
\vec{y}'|)F_{1}(|\vec{x}'-\vec{y}|)
\delta_{lj}(x_{k}' - y_{k})(x'_{n}-y_{n})
\nonumber\\
&+& F_{1}(|\vec{x}'- \vec{y}|)F_{1}(|\vec{x}-
\vec{y}'|)\delta_{kn}\delta_{jl}]. 
\label{B.6}
\end{eqnarray}
Recall that, for a generic function $f(r)$ (where $r=
|\vec{r}|$, $r^a = x^a - y^a$), the following trivial relation
holds
\begin{eqnarray}
\frac{\partial}{\partial x^{i}} \frac{\partial}{\partial y^{m}} f(r) =
- \delta_{im} \frac{1}{r} \frac{\partial f(r)}{\partial{r}} + \frac{r^{i}
r^{m}}{r^3} \frac{\partial f(r)}{\partial r} 
- \frac{r^{i}r^{m}}{r^2} \frac{\partial^2 f(r)}{\partial r^2}.
\nonumber
\end{eqnarray}
After having  performed the derivatives in Eq. (\ref{B.5}) we can
contract the various Levi-Civita tensors with the Kroeneker
symbols. By then taking the limits indicated in (\ref{B.3}) and
(\ref{B.5}) we obtain
\begin{eqnarray}
& & \biggl\langle\left(\vec{H}
\cdot\vec{\nabla}\times\vec{H}
\right)(\vec{x})
\left(\vec{H}\cdot\vec{\nabla}\times\vec{H}\right)
(\vec{x} +\vec{r}) \biggr\rangle = 
\nonumber\\
& & -\frac{4}{r}F_{1}(r) \frac{d F_{2}(r)}{dr} 
- 2 F_{1}(r) \frac{d^2 F_{1}(r)}{dr^2} 
+ 4 r^2 [F_{2}(r)]^2 + 2 r F_{1}(r) \frac{dF_{2}(r)}{d r}
\nonumber\\
& & - 6 r F_{2}(r) \frac{dF_{1}(r)}{dr} + 6 F_{1}(r) F_{2}(r) 
+ 2 \left(\frac{dF_{1}(r)}{dr} \right)^2.
\end{eqnarray} 
In this form the four-point correlation function is completely expressed in
terms of the two-point function. Of course we stress that this
decomposition holds provided the fields are stochastically
distributed, namely if and only if (\ref{B.4}) is satisfied.

\renewcommand{\theequation}{B.\arabic{equation}}
\setcounter{equation}{0}
\section{Magnetic helicity correlations ($\Gamma <\Gamma_{H}$)}

If $\Gamma<\Gamma_{H}$ the correlation function appearing in the final
expression of the level of the fluctuations turns out to be the
stochastic average of a quantity that contains the magnetic
helicity in the numerator and the hypermagnetic energy density in the
denominator. Even if this case turns out to be less relevant for the
phenomenological considerations presented in the main discussion, we
want to outline the main idea that can be used in order to get a
large-scale estimate of 
\begin{equation}
\biggl\langle\left(\frac{\vec{H}\cdot\vec{\nabla}\times\vec{H}}{H^2}
\right)(\vec{x}_{1})
\left(\frac{\vec{H}\cdot\vec{\nabla}\times\vec{H}}{H^2}\right)
(\vec{x}_{2}) \biggr\rangle.
\label{A.1}
\end{equation}
The strategy we will use will be to express (\ref{A.1}) in terms of an
appropriate path integral whose functional derivatives will reproduce
the correlation function we want to compute. From Eq. (\ref{A.1}) we
have, formally
\begin{eqnarray}
& &\biggl\langle\left(\frac{\vec{H}\cdot\vec{\nabla}\times\vec{H}}{H^2}\right)
(\vec{x}_{1})
\left(\frac{\vec{H}\cdot\vec{\nabla}\times\vec{H}}{H^2}\right)
(\vec{x}_{2}) \biggr\rangle=
\nonumber\\
& &\lim_{{\vec{x}}_{1}'\rightarrow {\vec{x}}_{1}} 
\lim_{{\vec{x}}_{2}'\rightarrow
{\vec{x}}_{2}}\lim_{\alpha\rightarrow 0} \lim_{\beta \rightarrow 0}
\epsilon_{ijk}\epsilon_{abc} \frac{\partial}{\partial x_{1}^i}
 \frac{\partial}{\partial x_{2}^a}\left(
\frac{\delta}{\delta J_{k}(\vec{x}'_{1})}
\frac{\delta}{\delta J_{j}(\vec{x}_{1})}\frac{\delta}
{\delta J_{c}(\vec{x}'_{2})}
\frac{\delta}{\delta J_{b}(\vec{x}_{2})} 
{\cal W}[\vec{J}]\right)_{\vec{J}=0}
\label{A.2}
\end{eqnarray}
where 
\begin{eqnarray}
& &{\cal W}[\vec{J}] = \frac{1}{16\pi^2} \int \frac{d^3 p}{p} \int
\frac{d^3 q}{q} \int D[\vec{H}]\times
\nonumber\\
& &\exp{}\big\{ -\frac{i}{2} \int d^3 x
\int d^3 y H_{i}(\vec{x})[K(\vec{x},\vec{y})]_{ij}
H_{j}(\vec{y}) + i \int J_{i}(\vec{x}) H_{i}(\vec{x})
d^3 x
\nonumber\\
&+&i p_{i} \int H_{i}(\vec{x})\delta^{(3)}(\vec{x} 
- \vec{x}'_{1}) d^3 x 
+iq_{j} \int
H_{j}(\vec{x}) \delta^{(3)}(\vec{x}- \vec{x}'_{2}) d^3x
\nonumber\\
&-&\alpha|\vec{p}|^2 -\beta|\vec{q}|^2\big\}.
\label{A.3}
\end{eqnarray}
In Eq. (\ref{A.3}) we used the fact that formally holds the
following relation
\begin{eqnarray}
\frac{1}{|{\vec{H}}|^2} = \frac{1}{4\pi}\lim_{\alpha\rightarrow 0}
\int \frac{d^3 p}{p} e^{i\vec{H}\cdot \vec{p} 
- \alpha |\vec{p}|^2}
\nonumber
\end{eqnarray}
By appropriately redefining the source term in the path integral,
 Eq.  (\ref{A.3}) can also be written as :
\begin{eqnarray}
&&{\cal W}[\vec{J}]= \frac{1}{16 \pi^2} 
\int\frac{d^3 p}{p} \int\frac{ d^3 q}{q} \int 
D[\vec{H}]\times
\nonumber\\
&&\exp{}\biggl\{ -\frac{i}{2} \int d^3 x\int d^3 y
H_{i}(\vec{x}) [K(\vec{x},\vec{y})]_{ij}
H_{j}(\vec{y}) -\alpha |\vec{p}|^2 - \beta|\vec{q}|^2
\nonumber\\
&&+i \int S_{i}(\vec{x}) H_{i}(\vec{x}) \biggr\} d^3x
\label{A.4}
\end{eqnarray}
where, in the present case, 
\begin{equation}
S_{i}(\vec{x})= J_{i}(\vec{x})+  p_{i}
\delta^{(3)}(\vec{x}- \vec{x}'_{1})+  q_{i} \delta^{(3)}
(\vec{x}- \vec{x}'_2).
\label{A.5}
\end{equation}
By defining 
\begin{eqnarray} 
h_{i}(\vec{x}) = \int d^3 y
[K^{\frac{1}{2}}(\vec{x},\vec{y})]_{ij} H_{j}(\vec{y})
\nonumber
\end{eqnarray}
we obtain that the argument of the exponential can be expressed as
\begin{equation}
\frac{i}{2}\int d^3 x \biggl\{ h_{i}(\vec{x}) - \int d^3 x
S_{i}(\vec{y})[K^{-\frac{1}{2}}(\vec{x},\vec{y})]_{ij}\biggr\}^2
-\frac{i}{2}\int d^3 x\int d^3 y
S_{i}(\vec{x})G_{ij}(|\vec{x}-\vec{y}|)S_{j}(\vec{y}).
\label{A.6}
\end{equation}

Notice that the symbol $ [K(\vec{x},\vec{y})]_{ij}$ must have the
following properties, which will be important also for the calculation
of the functional integral:
\begin{eqnarray}
&\int &d^3 y  [K^{\frac{1}{2}}(\vec{x},\vec{y})]_{ij} 
[K^{\frac{1}{2}}(\vec{y},\vec{z})]_{jk} = 
 [K(\vec{x},\vec{z})]_{ik}
\nonumber\\
&\int & d^3 y  [K^{\frac{1}{2}}(\vec{x},\vec{y})]_{ij} 
 [K^{-\frac{1}{2}}(\vec{y},\vec{z})]_{jk} = \delta_{ik} 
\delta^{(3)}(\vec{x}-\vec{z}) 
\nonumber\\
&\int & d^3 y  [K^{-\frac{1}{2}}(\vec{x},\vec{y})]_{ij} 
 [K^{-\frac{1}{2}}(\vec{y},\vec{z})]_{jk} 
= [K(\vec{x},\vec{z})^{-1}]_{ik}
\nonumber\\
& &[K^{-1}(\vec{x},\vec{y})]_{ij} =
-G_{ij}(|\vec{x}-\vec{y}|).
\label{A.7}
\end{eqnarray}

We can integrate the part that is quadratic in the fields; then
${\cal W}[\overline{J}]$ becomes
\begin{eqnarray}
{\cal W}[\overline{J}] &=& W[0] \int\frac{d^3 p}{p} \int \frac{d^3 q}{q} 
\exp{}\{-\frac{i}{2} \int d^3 x \int d^3 y S_{i}(\overline{x})
G_{ij}(|\overline{x}-\overline{y}|)S_{j}(\overline{y})
\nonumber\\
&-& \alpha|\overline{p}|^2 - \beta |\overline{q}|^2
\},
\label{A.8}
\end{eqnarray}
where ${\cal W}[0]$ is the usual Jacobian.
Using the definition of $S_{i}(\vec{x})$ we obtain for ${\cal
W}[\overline{J}]$:
\begin{eqnarray}
& &\frac{{\cal W}[\vec{J}]}{{\cal W}[0]}= \int\frac{d^3 p}{p} \int
\frac{d^3 q}{q} e^{C(q_{i}, p_{i};q_{j},p_{j})} 
\nonumber\\
&\times & \exp{}\biggl\{-\frac{1}{2}
\int d^3 x\int d^3 y J_{i}(\vec{x})
G_{ij}(|\vec{x}-\vec{y}|) J_{j}(\vec{y})- i \int d^{3} x
J_{l}(x) L_{l}(x) \biggr\}
\label{A.9}
\end{eqnarray}
with
\begin{equation}
L_{l}(x) = p_{m}G_{ml}(|\vec{x} -\vec{x}'_{1}|) 
+ q_{m}G_{ml}(|\vec{x} -\vec{x}'_{2}|)
\label{A.10}
\end{equation}
and
\begin{equation}
C(q_{i},p_{i};q_{j}, p_{j})  = -\frac{1}{2}q_{i}G_{ij}(0)q_{j} -
 \frac{1}{2}p_{i}G_{ij}(0)p_{j} 
- q_{i} G_{ij}(|\vec{x}'_1 - \vec{x}'_{2}|)p_{j}
- \alpha|\vec{p}|^2 - \beta |\vec{q}|^2
\label{A.11}
\end{equation}
($G_{ij}(0)$ comes because there are two delta functions centred at
the same point for the terms quadratic in $q_{i}$ and
$p_{i}$).
Performing the  functional derivatives we obtain (evaluating
the generating function for $J=0$)
\begin{eqnarray}
& &\left( \frac{\delta}{\delta J_{k}(\vec{x}'_{1})}
\frac{\delta}{\delta J_{j}(\vec{x}_{1})}\frac{\delta}
{\delta J_{c}(\vec{x}'_{2})}
\frac{\delta}{\delta J_{b}(\vec{x}_{2})}
W[\vec{J}]\right)_{\vec{J}=0}=
\nonumber\\
& & \{[G_{kj}(|\vec{x}_{1}
-\vec{x}'_{1}|)G_{cb}(|\vec{x}_{2}-\vec{x}'_{2}|)
\nonumber\\
&+& G_{kc}(|\vec{x}'_{1}
-\vec{x}'_{2}|)G_{jb}(|\vec{x}_{1}-\vec{x}_{2}|)
+G_{kb}(|\vec{x}'_{1}
-\vec{x}_{2}|)G_{jc}(|\vec{x}_{1}-\vec{x}'_{2}|)]
\nonumber\\
&- & L_{b}(\vec{x}_{2})L_{c}(\vec{x}'_{2})
G_{jk}(|\vec{x}_{1}-\vec{x}'_{1}|)
-L_{k}(\vec{x}'_{1})L_{j}(\vec{x}_{1})
G_{bc}(|\vec{x}_{2}-\vec{x}'_{2}|)
\nonumber\\
&-& L_{c}(\vec{x}'_{2})L_{k}(\vec{x}'_{1})
G_{bj}(|\vec{x}_{1}-\vec{x}_{2}|)
- L_{b}(\vec{x}_{2})L_{k}(\vec{x}'_{1})
G_{cj}(|\vec{x}_{1}-\vec{x}'_{2}|)
\nonumber\\
&-& L_{c}(\vec{x}'_{2})L_{j}(\vec{x}_{1})
G_{bk}(|\vec{x}'_{1}-\vec{x}_{2}|)
- L_{b}(\vec{x}_{2})L_{k}(\vec{x}'_{1})
G_{kc}(|\vec{x}'_{1}-\vec{x}'_{2}|)
\nonumber\\
&+& L_{k}(\vec{x}'_{1})
L_{j}(\vec{x}_{1})L_{b}(\vec{x}'_{2})L_{c}(\vec{x}_{2})\}{\cal
W}[0].
\label{A.12}
\end{eqnarray}
Notice that the fifth and sixth terms of Eq. (\ref{A.12}) vanish when
contracted with the epsilon tensors.
In order to perform the integration over $p$ and $q$, we have to expand
the expression giving us the correlation function for
\begin{equation}
g(r) =\frac{{\cal G}(R)}{{\cal G}(0)} < 1,
\label{A.13}
\end{equation}
(see also Eq. (\ref{4.10})). 
This approximation holds for sufficiently large scales, provided the
Green functions decay for $R \gg 1$. This requirement is
automatically satisfied in our case, since we only consider the
situation where the energy spectrum is increasing in frequency
(i.e. blue or violet spectra).
We now take the limits for $\vec{x}_{1}'\rightarrow
\vec{x}_{1}$, and for  $\vec{x}_{2}'\rightarrow
\vec{x}_{2}$ and Eq. (\ref{A.2}) becomes:
\begin{eqnarray}
& &
\biggl\langle\left(\frac{\vec{H}\cdot\vec{\nabla}\times\vec{H}}{H^2}
\right)(\vec{x})
\left(\frac{\vec{H}\cdot\vec{\nabla}\times\vec{H}}{H^2}\right)
(\vec{x}+\vec{r}) \biggr\rangle 
\nonumber\\
&\simeq& \lim_{\alpha\rightarrow 0}\lim_{\beta \rightarrow 0}\frac{1}{16\pi^2}
\int \frac{d^3 p}{p} \int \frac{d^3 q}{q}[ A_{1}(r) + q^2 A_{2}(r) -
p^2 A_{3}(r) 
\nonumber\\
&+& 2 A_{4}(r) (\vec{q}\cdot\vec{p})] \exp{\big\{ -
\frac{\langle H^2(\vec{x})\rangle}{2} ( q^2 + p^2) -
\alpha|\vec{q}|^2 - \beta|\vec{p}|^2 + O(g(r)) \big\}},
\label{A.16}
\end{eqnarray}
where
\begin{eqnarray}
A_{1}(r) &=& \biggl\langle\left( (\vec{H}\cdot\vec{\nabla}\times
\vec{H})(\vec{x})(
\vec{H}\cdot\vec{\nabla}\times\vec{H})
(\vec{x}+\vec{r})\right)\biggr\rangle
\nonumber\\
A_{2}(r) &=& A_{3}(r)= 
\frac{1}{2} [\langle H^2(\vec{x})\rangle]^3 g(r)
\nonumber\\
A_{4}(r) &=& \frac{1}{16} [\langle H^2(\vec{x})\rangle ]^3 g(r).
\label{A.17}
\end{eqnarray}
We can now integrate over $q$ and $p$. We notice that since the
integral in convergent also for $\alpha, \beta\rightarrow 0$ the
limits can be  taken  before the integration.
It is convenient to perform the integration over $q$ and $p$
separately; in this way, after
angular integration, the apparently Gaussian integrals can  be
expressed as ordinary exponential integrals of the type
\begin{eqnarray}
\int_{0}^{\infty} \lambda^{n} e^{- a \lambda} = \frac{\Gamma(n +
1)}{a^{n +1}}.
\end{eqnarray} 
After integration, $A_{2}(r)$ and $A_{3}(r)$ cancel whereas the contribution
of the term containing $A_{4}(r)$ vanishes because of the angular integration.
The final result obtained in the assumption that the Green functions
decay at large distances is then
\begin{eqnarray}
 \biggl\langle\left(\frac{\vec{H}\cdot\vec{\nabla}\times\vec{H}}{H^2}
\right)(\vec{x})
\left(\frac{\vec{H}\cdot\vec{\nabla}\times\vec{H}}{H^2}\right)
(\vec{x} +\vec{r}) \biggr\rangle \simeq
\nonumber\\
\frac{\langle(\vec{H}\cdot
\vec{\nabla}\times\vec{H})(\vec{x})(
\vec{H}\cdot\vec{\nabla}\times\vec{H})
(\vec{x}+\vec{r})\rangle}{ \langle H^2(\vec{x}) \rangle^2} 
+ O(g(r)),
\label{A.19}
\end{eqnarray}
which is exactly what we report in Section 4.
The method used in the present Appendix can also be exploited in order
to compute further corrections, if needed.
\end{appendix}

\end{document}